\documentclass[
  twocolumn
]{aastex631}

\makeatletter
\let\frontmatter@title@above=\relax
\makeatother

\usepackage{chemformula}

\usepackage{CJK}

\newcommand\nocolor{\multicolumn2c{\nodata}}

\newcommand\mps{m~s$^{-1}$}

\newcommand\kgps{kg~s$^{-1}$}

\newcommand\coo{\mbox{CO$_2$}}
\newcommand\water{\mbox{H$_2$O}}

\newcommand\afrho[1][\theta]{\mbox{$A(#1)f\rho$}}
\newcommand\afr{\mbox{$Af\rho$}}

\newcommand\rh{\mbox{$r_{\mathrm{h}}$}}
\newcommand\inv[2][1]{$\textrm{#2}^{-#1}$}

\newcommand\specgrad[1]{$#1$\% per 100 nm}
\newcommand\SV[1]{$S_V$=\specgrad{#1}}

\newcommand\sst{{Spitzer Space Telescope}}

\newcommand\di{{Deep Impact}}

\graphicspath{{./}{figures/}}

\accepted{2025 June 15}
\submitjournal{The Planetary Science Journal}

\shorttitle{Comet 243P/NEAT}
\shortauthors{Kelley et al.}

\begin{document}
\begin{CJK*}{UTF8}{gbsn}

  \title{A Large Outburst, Coma Asymmetries, and the Color of Comet 243P/NEAT}

  \author[0000-0002-6702-7676]{Michael S. P. Kelley}
  \altaffiliation{Visiting Astronomer at the Infrared Telescope Facility, which is operated by the University of Hawaii under contract NNH14CK55B with the National Aeronautics and Space Administration.}
  \affiliation{Department of Astronomy, University of Maryland, College Park, MD 20742-0001, USA}
  \email{msk@astro.umd.edu}

  \author[0000-0001-8541-8550]{Silvia Protopapa}
  \altaffiliation{Visiting Astronomer at the Infrared Telescope Facility, which is operated by the University of Hawaii under contract NNH14CK55B with the National Aeronautics and Space Administration.}
  \affiliation{Southwest Research Institute, Boulder, CO 80302, USA}

  \author[0000-0002-2668-7248]{Dennis Bodewits}
  \affiliation{Physics Department, Edmund C. Leach Science Center, Auburn University, Auburn, AL 36832, USA}

  \author{Aren N. Heinze}
  \affiliation{DiRAC Institute and the Department of Astronomy, University of Washington, 3910 15th Avenue NE, Seattle, WA 98195 USA; aheinze@uw.edu}
  \affil{Institute for Astronomy, University of Hawaii, Honolulu, HI 96822-1839, USA}

  \author[0000-0001-9784-6886]{Youssef Moulane}
  \affiliation{Physics Department, Edmund C. Leach Science Center, Auburn University, Auburn, AL 36832, USA}
  \affiliation{School of Applied and Engineering Physics, Mohammed VI Polytechnic University, Benguerir 43150, Morocco.}

  \author[0000-0002-4838-7676]{Quanzhi Ye (叶泉志)}
  \affiliation{Division of Physics, Mathematics and Astronomy, California Institute of Technology, Pasadena, CA 91125, U.S.A.}
  \affiliation{IPAC, California Institute of Technology, 1200 E. California Blvd, Pasadena, CA 91125, USA}
  \affiliation{Department of Astronomy, University of Maryland, College Park, MD 20742-0001, USA}
  \affiliation{Center for Space Physics, Boston University, 725 Commonwealth Ave, Boston, MA 02215, USA}

  \author[0000-0002-4950-6323]{Bryce Bolin}
  \altaffiliation{NASA Postdoctoral Program Fellow}
  \affiliation{Goddard Space Flight Center, 8800 Greenbelt Road, Greenbelt, MD 20771, U.S.A.}

  \author[0000-0002-3657-4191]{Simon Conseil}
  \affiliation{Universit\'e Claude Bernard Lyon 1, CNRS/IN2P3, IP2I Lyon, UMR 5822, Villeurbanne, F-69100, France}

  \author{Tony L. Farnham}
  \affil{Department of Astronomy, University of Maryland, College Park, MD 20742-0001, USA}

  \author{Lori Feaga}
  \affil{Department of Astronomy, University of Maryland, College Park, MD 20742-0001, USA}

  \author{Xing Gao (高兴)}
  \affiliation{Xinjiang Astronomical Observatory, Urumqi, Xinjiang 830011, China}

  \author[0000-0003-2549-3326]{Chih-Hao Hsia (夏志浩)}
  \affiliation{The Laboratory for Space Research, Faculty of Science, The University of Hong Kong, Cyberport 4, Hong Kong, China}

  \author{Emmanuel Jehin}
  \affiliation{Space sciences, Technologies \& Astrophysics Research (STAR) Institute, University of Li\`ege}

  \author[0000-0001-5390-8563]{Shrinivas R. Kulkarni}
  \affiliation{Division of Physics, Mathematics, and Astronomy, California Institute of Technology, Pasadena, CA 91125, USA}

  \author[0000-0003-2451-5482]{Russ R. Laher (\begin{CJK*}{UTF8}{bsmi}良主嶺亞\end{CJK*})}
  \affiliation{IPAC, California Institute of Technology, 1200 E. California Blvd, Pasadena, CA 91125, USA}

  \author[0000-0002-3818-7769]{Tim Lister}
  \affiliation{Las Cumbres Observatory (LCOGT), 6740 Cortona Drive Suite 102, Goleta, CA 93317, USA.}

  \author[0000-0002-8532-9395]{Frank J. Masci}
  \affiliation{IPAC, California Institute of Technology, 1200 E. California Blvd, Pasadena, CA 91125, USA}

  \author[0000-0003-1227-3738]{Josiah Purdum}
  \affiliation{Caltech Optical Observatories, California Institute of Technology, Pasadena, CA 91125, USA}

  \author[0000-0002-5033-9593]{Bin Yang (杨彬) }
  \affiliation{Instituto de Estudios Astrof\'isicos, Facultad de Ingenier\'ia y Ciencias, Universidad Diego Portales, Av. Ej\'ercito 441, Santiago, Chile}
  \affiliation{Planetary Science Institute, 1700 E. Fort Lowell Rd., Ste. 106 Tucson, AZ 85719-2395, U.S.A.}

  \begin{abstract}
    Water ice is a fundamental building material of comets and other bodies in the outer solar system.  Yet, the properties of cometary water ice are challenging to study, due to its volatility and the typical distances at which comets are observed.  Cometary outbursts, impulsive mass-loss events that can liberate large amounts of material, offer opportunities to directly observe and characterize cometary water ice.  We present a study of comet 243P/NEAT, instigated by a $-3$~mag outburst that occurred in December 2018.  Optical images and a 251-day lightcurve were examined to characterize the outburst and the comet's quiescent activity.  Variations in the quiescent lightcurve appear to be dominated by coma asymmetries, rather than changing activity levels as the comet approached and receded from the Sun.  Furthermore, the lightcurve shows evidence for 1 to 2 additional small outbursts (--0.3~mag) occurring in September 2018.  The large December 2018 outburst likely ejected water ice grains, yet no signatures of ice were found in color photometry, a color map, nor a near-infrared spectrum.  We discuss possible dynamical and thermal reasons for this non-detection.  In this context, we examined the comae of comets 103P/Hartley 2 and C/2013 US$_{10}$ (Catalina), and show that a one-to-one mapping between continuum color and the presence of water ice cannot be supported.  We also discuss possible causes for the large outburst, and find that there is an apparent grouping in the kinetic energy per mass estimates for the outbursts of 5 comets.
  \end{abstract}


  \section{Introduction} \label{sec:intro}
  Cometary outbursts are short-lived mass-loss events.  The possible mechanisms of outbursts are varied, including rotational breakup, land slides, cliff collapse, and explosive out-gassing, the latter potentially from ice phase transitions or sub-surface gas pressure build-up \citep[e.g.,][]{pajola17, agarwal17-outburst, jewitt20-borisov, noonan21-outbursts}.  These events can provide opportunities to study materials from particular regions on the nucleus, especially sub-surface locations previously isolated from intense solar heating.  Thus, the properties of the material released by outbursts are not necessarily the same as those of comae in quiescence.  By comparing the coma composition before, during, and after the event, we may probe the nature of the outburst and the heterogeneity of the nucleus.  However, obtaining pre-outburst coma compositions for all comets that might outburst is presently infeasible, and post-outburst, quiescent coma studies may be limited by the intrinsic faintness of the target or other observational constraints.  This issue is partially mitigated by observations by wide-field sky surveys, but quiescent spectroscopic data are not routinely obtained.  Yet scientific gains may still be achieved by comparing all outbursts to each other and to the general properties of comets.

  In particular, there is the potential to directly study cometary water ice through observations of outbursts.  Water ice is a major component of cometary nuclei, and the physical properties of the ice (abundance, grain size, purity, chemical phase) affects the comet's mass-loss rate, the propagation of thermal energy into the interior, and the storage and release of minor volatile species \citep{prialnik04}.  Sub-surface excavation by an outburst could eject large amounts of water ice into the coma, making it available for study.  Though observations may provide indirect evidence of water ice \citep{ahearn84-bowell, villanueva11-boattini, bonev13, bodewits14-garradd}, the best way to discern its physical properties in a cometary coma is directly through spectroscopic means.

  Water ice has been telescopically observed in three outbursts through near-infrared spectroscopy: the mega-outburst of comet 17P/Holmes in 2007 \citep{yang09-holmes}, the outburst of comet P/2010 H2 (Vales) \citep{yang10-iauc}, and a medium-large outburst of comet 29P/Schwassmann-Wachmann~1 \citep{protopapa21-atel14961}.\footnote{Water ice has also been observed in the ultra-violet in a small outburst of comet 67P/Churyumov-Gerasimenko \citep{agarwal17-outburst}.}   However, not all outbursts exhibit the near-infrared spectral signatures of water ice, due to signal-to-noise limitations and/or low water ice abundances \citep{moretto17-tempel1, bockelee-morvan17-outbursts}.  In cases of a significant non-detection, it is critical to consider the sublimation timescale of water-ice grains \citep[e.g.,][]{protopapa18-catalina}, as they may sublimate to extinction before obtaining follow-up observations.  In fact, near-infrared spectroscopy of outbursts may improve our knowledge of water ice grain sublimation; knowing the epoch of ejection from the nucleus informs us on the sublimation lifetime of the material observed.

  An outburst of comet 243P/NEAT was discovered by the Asteroid Terrestrial-Impact Last Alert System (ATLAS) as a brightening of at least --2.5~mag between 2018 December 10 08:50 and December 12 08:08 UTC \citep{heinze18-cbet4587}.  Comet 243P/NEAT was previously discovered by the Near-Earth Asteroid Tracking (NEAT) survey in September 2003 and given the provisional designation P/2003~S2 \citep{hicks03-iauc8209}.  It is a member of the Jupiter-family dynamical class with a 7.5-yr orbital period, and passed the Sun at a perihelion distance of 2.45~au on 2018 August 26 00:10 UTC (Minor Planet Circular 111774).  \citet{reach07} observed the comet at 3.6~au post-perihelion with the \sst{} in the course of their survey of cometary dust trails, and found it to have a long and narrow dust trail near the projected orbit of the comet, indicating the presence of 100~\micron{} and larger dust grains.  \citet{fernandez13} attempted to observe thermal emission from the nucleus with the \sst{} at 5.2~au, but did not detect the comet, and provided a 3$\sigma$ upper-limit estimate of 0.6~km for the radius.  \citet{mazzottaepifani08} detected the comet as a point source at a heliocentric distance, \rh, of 4.0~au in $R$-band images, and estimated a nucleus radius of 0.8--1.55~km, assuming a geometric albedo of 0.04.  Their minimum value is based on a correction for potentially unresolved coma.  Both the \citet{fernandez13} and \citet{mazzottaepifani08} results can be simultaneously true if the comet has an axial ratio of $a/b>1.3$ ($3\sigma$ lower limit), a modest value for cometary nuclei \citep{knight24-comets3}.

  After receiving notice of the outburst we initiated a follow-up study to characterize the comet and this event.  The first targeted data were taken within 3 days of the outburst discovery image.  In this paper, we present photometric, morphological, and spectroscopic studies of these data, and additional data identified in automated sky survey archives.  We characterize the dynamics of the quiescent comet and the December 2018 outburst.  We also test for the presence of water ice grains in the outburst ejecta.  For context, we analyze observations of the icy comae of comets 103P/Hartley~2 and C/2013 US$_{10}$ (Catalina).

  \section{Observations} \label{sec:observations}
  \subsection{Optical Imaging}
  Optical images of comet 243P were taken at several observatories, both synoptically as part of all-sky surveys, and through targeted observations.  Survey observations were taken with the Zwicky Transient Facility (ZTF), using the 1.2-m Samuel Oschin Telescope at Palomar Observatory, two ATLAS 0.5-m telescopes at the Mauna Loa and Haleakala Observatories in Hawaii, and the Catalina Sky Survey's 0.7-m Schmidt and 1.5-m Cassegrain telescopes on Mt.\ Bigelow.  Targeted observations were obtained with the 3.5-m Astrophysical Consortium (ARC) telescope (Apache Point Observatory, New Mexico), the 5-m Hale Telescope (Palomar Observatory), the Las Cumbres Observatory Global Telescope (LCOGT) network of 1-m telescopes (Haleakala Observatory, Hawaii; McDonald Observatory, Texas; South African Astronomical Observatory, South Africa; Cerro Tololo International Observatory, Chile; Siding Spring Observatory, Australia), the 4.3-m Lowell Discovery Telescope (LDT; Lowell Observatory, Arizona),
  the Lulin One-meter Telescope (LOT; Lulin Observatory, Taiwan), the 0.6-m Ningbo beureau of Education and Xinjiang observatory Telescope (NEXT; Xingming Observatory, China), and the 0.6-m TRAnsiting Planets and PlanetesImals Small Telescopes (TRAPPIST; Oukaimeden Observatory, Morocco; European Southern Observatory, La Silla, Chile).

  The data considered here span the 251 day period from 2018 June 20 to 2019 February 26, around the comet's perihelion ($T_P=$ 2018 August 26 00:10 UTC at 2.45~au).  The heliocentric distance ranged from 2.49~au pre-perihelion to 2.72~au post-perihelion.  The phase angles and geocentric distances varied from 5\degr{} to 24\degr, and 1.50 to 2.97~au, with the minima occurring on 2018 October 15, 50.50~days after perihelion.  Broad-band filters were used at all telescopes (Johnson-Cousins, SDSS, ZTF, and ATLAS filter sets), except for the Catalina Sky Survey data and a single epoch with a Las Cumbres Observatory telescope, which were unfiltered.  Data are primarily taken through SDSS-like filters, or through the ATLAS $c$- and $o$-filters (effectively $g+r$ and $r+i$, respectively).  However, some data sources used Johnson-Cousins $B$, $V$, $R$, and/or $I$ filters.  In addition, the LDT observations include data taken through the $BC$ and $RC$ narrow-band filters from the HB filter set \citep{farnham00}.

  Details on the instruments, observations, and data reduction steps are presented in the individual data summaries in Appendix~\ref{app:obs}.  An observing log is also presented in the Appendix (Table~\ref{tab:obs}).  Observations of the comet were avoided during a period of low Lunar elongation (minimum 2\degr{} on 2018 Dec 18).  The survey telescopes (ZTF, ATLAS, and CSS) tracked the sky at the sidereal rates.  The comet's non-sidereal motion was $\leq57$\arcsec~\inv{hr}, or $\leq0.5\arcsec$ ($\leq0.5$~pix) in all cases.  All other telescopes tracked the comet based on its ephemeris, unless otherwise noted.  All images were bias, dark (when applicable), and flat-field corrected with contemporaneously taken calibration data.  The data were photometrically calibrated to the Pan-STARSS~1 photometric system \citep{tonry12-ps1}.  Aperture photometry for most observations were measured within a 10,000-km radius (4\farcs6 to 9\farcs2) centered on the nucleus. No significant difference between the sidereally tracked and non-sidereally tracked data is apparent in the resulting photometry.  Table~\ref{tab:obs} includes a measurement of the image quality, derived from the radial profiles of background sources in the comet data or with separate sidereally tracked images, unless otherwise noted.

  \subsection{Near-Infrared Spectroscopy}\label{sec:irtf}
  We also observed comet 243P with the SpeX instrument at the 3.2-m NASA Infrared Telescope Facility (IRTF) (Mauna Kea Observatory, Hawaii).  SpeX is a medium-resolution near-infrared spectrograph \citep{rayner03}.  We observed the comet with the low-resolution mode ($0\farcs8\times60\arcsec$ slit for $R=\lambda/\Delta\lambda\sim75$), covering 0.7--2.5~\micron.  At the mean time of the observations, 2018 December 15 at 05:51 UTC, the comet was 2.56~au from the Sun, and 1.91~au from the Earth.  The mean airmass was 1.06 (range 1.05--1.07), and seeing was 0\farcs6, based on J-band images of a star with the guiding camera.  The raw data are available from the IRTF data archive under program 2018B118.  Twelve 120~s exposures were taken, with the comet nodded between two positions along the slit separated by 30\arcsec.

  The data were reduced using IDL codes based on Spextool \citep{cushing04}. Spextool was designed to reduce data taken with the 15\arcsec-long slit and does not support our observations taken with the 60\arcsec-long slit.  These data reduction tools were previously implemented and tested against observations of comets C/2013 US10 and 46P/Wirtanen \citep{protopapa18-catalina, protopapa21-wirtanen}, ensuring their reliability for our analysis. Spectra were extracted, wavelength-calibrated, and combined using a robust mean with a 2.5$\sigma$-clipping threshold.  The background was measured by nodding the comet parallel to the long slit, and the data were flat-fielded and wavelength calibrated with SpeX's internal calibration lamps.  Spectra were extracted with a 1\arcsec{} radius aperture, and calibrated to relative reflectance with solar-type (G2V) star HD 1368 \citep{houk99}. The aperture size was chosen to maximize the signal from the comet without adding undue noise.  Tests with a 3\arcsec-wide aperture produced results consistent with our adopted 1\arcsec-wide aperture.  The star was observed immediately after the internal calibration data at a mean airmass of 1.11 (range 1.10 to 1.11) with ten 5~s exposures, and the same nod positions as the comet.  Observations of the comet and standard were taken with the slit aligned along the parallactic angle.

  Following \citet{protopapa21-wirtanen}, we compared spectra of three standard stars taken throughout the night by measuring the continuum slope of the ratio of paired stars (HD 1368 and SA 93 101 from program 2018B118 and HD 30246 from 2018B006).  The slopes were 0.1 and 0.3\% per 100~nm for the ratios HD 1368/HD 30246 and SA 93 1010/HD 30246, respectively.  Given the low-number statistics, we take the maximum value, 0.3\% per 100~nm, as the 1$\sigma$ slope uncertainty.  This is the same uncertainty adopted by \citet{protopapa21-wirtanen} in their analysis of six solar analogs observed with SpeX.

  \subsection{103P/Hartley 2}\label{sec:deepimpact}
  As a point of reference for the 243P data set, we also investigated the coma of comet 103P/Hartley 2, as imaged and spectrally mapped by the Deep Impact flyby spacecraft's High Resolution Instrument (HRI) and Medium Resolution Instrument (MRI).  Details on the Deep Impact flyby, its instruments and their calibrations are available from \citet{ahearn11}, \citet{hampton05}, and \citet{klaasen08,klaasen13}.  The MRI had a 12-cm primary mirror and a single instrument, a 1024$\times$1024 split-frame transfer CCD, with a 10~\textmu{}rad pixel scale.  The HRI had a 30-cm primary, and the optical path fed a near-infrared spectrometer (HRI-IR) and a second CCD with a 2~\textmu{}rad pixel scale (HRI-Vis).  HRI-IR utilized a double-prism disperser, with a spectral resolving power of 200 to 700 covering 1.05 to 4.8~\micron, and an angular pixel size of 5~\textmu{}rad.  The entrance aperture to the spectrometer was a long-slit with a width of 10~\textmu{}rad, and most data, including the observations analyzed here, were taken in 2$\times$2 pixel binned mode.  The HRI was significantly out of focus, with a PSF FWHM of about 9~pix in the optical \citep{klaasen08}.

  \citet{protopapa14} analyzed two spectral maps of comet Hartley 2 taken with the HIR-IR spectrometer, showing the resolved nucleus, and an inner coma of dust, ice, and gas.  The maps were taken 540 and 1380~s after the closest approach to the comet at distances of 5478 and 17,295~km, respectively.  Therefore, the first of the two data sets has a finer spatial sampling, 110~m~pix$^{-1}$ (binned).  We analyze this data set (observation 5006000) as a point of comparison for our near-infrared spectrum of comet 243P.  The original spectral maps as used by \citet{protopapa14} were obtained.  A row-by-row background subtraction was applied to the data, derived from the sigma-clipped median of the 40 most distant pixels from the comet \citep[a similar approach as by][]{protopapa14}.  This correction was most significant at the short wavelength end of the spectrum.  In addition, column-by-column variations were mitigated by high-pass filtering, by removing the column-by-column median difference between each pixel and the median of its neighbors in the same row (5~pix window).  Spectra of regions of interest were extracted from the resulting data cube by averaging.  Single-point spectral outliers were identified by comparison to the local median, and removed as needed (e.g., if they deviate by more than 2.5$\sigma$).  Our work will be focused on the continuum spectrum, and this outlier rejection may affect the shapes of gas emission bands, but is sufficient for the study of broad continuum features.  The spectra are converted into reflectance values by normalization with the CALSPEC model solar spectrum \citep{bohlin14-calspec}.

  We searched the HRI-Vis and MRI calibrated data archives \citep{mclaughlin11-dixi-hri,mclaughlin11-dixi-mri} for images taken close in time to our chosen post-closest approach spectral map.  For this investigation, we limited our analysis to ``blue,'' ``green,'' and ``red'' filter pairs to approximately span the optical data in our ground-based investigation of comet 243P.  Two color pairs (4 images total) were identified and are summarized in Table~\ref{tab:di}.  The selected data have central wavelengths of 454, 526, and 744~nm.  Compare these to our 243P data, which are primarily through SDSS-like $g'$ and $r'$ filters ($\sim$480 and $\sim$630 nm), or Johnson-Cousins $B$, $V$, $R$, and $I$ filters, with central wavelengths spanning 440 to 810~nm.  The Deep Impact filters have the potential for gas contamination.  Although the narrowband 526~nm filter was designed to avoid gas emission, the broader 454 and 744 filters cover emission bands from, e.g., \ch{C3} (410~nm), \ch{C2} (470~nm), \ch{NH2} (740 nm), and CN (420 and 790 nm) \citep{cochran12,kwon17-catalina}.  We assume the filters are continuum dominated and do not correct for any gas contamination.  The selected images were previously aligned (including corrections for the changing spacecraft-comet distance) and analyzed by \citet{li13-hartley2}, and we use their results in our work.

  \begin{deluxetable*}{rccccc}
    \tablecaption{Deep Impact Optical Images \label{tab:di}}
    \tablehead{
      \multicolumn{1}{c}{Image}
      & Date
      & $\Delta$
      & Filter
      & $\lambda_{\mathrm{eff}}$
      & $\Delta\lambda$ \\
      & (UTC)
      & (km)
      &
      & (nm)
      & (nm)
    }
    \startdata
    hv10110414\_5006048 & 2010 Nov 04 14:09:04 & 6892 & Blue & 454 & 100 \\
    mv10110414\_5006060 & 2010 Nov 04 14:09:11 & 6982 & GC & 526 & 5.6 \\
    mv10110414\_5006060 & 2010 Nov 04 14:09:19 & 7071 & Red & 744 & 100 \\
    hv10110414\_5006052 & 2010 Nov 04 14:09:38 & 7310 & Red & 744 & 100 \\
    \enddata
    \tablenotetext{}{Columns| Date: observation mid time; $\Delta$: spacecraft-comet distance; Filter: Deep Impact filter name (GC = green continuum); $\lambda_\mathrm{eff}$: filter effective wavelength for solar radiation; $\Delta\lambda$: filter bandwidth.}
  \end{deluxetable*}

  \section{Dust Dynamical Model} \label{sec:dynamical-model}
  Interpretation of coma and outburst morphologies may be made with a dust dynamical model and simulated images.  We use the 3D Monte Carlo model of \citet{kelley23-pycometsuite-v1.0.0} with heritage from \citet{kelley06-phd} and \citet{kelley09-cg}.  The version of the dynamical model used for this work solves equations of motion with the Bulirsch-Stoer algorithm of \citet{bader83-bsint} in the GNU Scientific Library.  Perturbations of the planets were not considered, only the Sun's gravity and radiation forces.  The free parameters of the model are production rate  ($Q$) as a function of time and/or heliocentric distance, ejection velocity (speed $s$, direction $\hat{v}$) as a function of grain size and heliocentric distance, dust size distribution, and dust composition.  Grain dynamics are accounted for by the dimensionless radiation pressure efficiency factor $\beta=F_r/F_g$, where $F_r$ is the force from solar radiation pressure, and $F_g$ is the force from solar gravity.  The ratio reduces to:
  \begin{equation}
    \beta = Q_{pr} \left(\frac{0.57~\micron{}}{a}\right)\left( \frac{1000 \mathrm{~kg~m}^{-3}}{\rho_g}\right),
  \end{equation}
  where $Q_{pr}$ is the size and composition dependent radiation pressure efficiency, $a$ is grain radius, and $\rho_g$ is the grain density \citep{burns79}.

  For all simulations, we used solid amorphous carbon grains with optical constants from \citet{edoh83} and a bulk density of 1.5~g~cm$^{-3}$ \citep{williams72-carbon}.  Grain radii ranged from $a_\mathrm{min}$=0.1~\micron{} to $a_\mathrm{max}$=1 cm, and radiation pressure efficiency factors are computed assuming solid spheres and Mie theory.\footnote{Modified \citealt{bohren83} code from B.~Draine, available at https://www.astro.princeton.edu/~draine/scattering.html.}  Dust ejection speeds are quoted for 1~\micron{} grains ($s_1$), and scaled by $a^{-1/2}$, where $a$ is the grain radius.  Here, ejection speed is the speed imparted on the grain by the time it decouples from the near-nucleus gas flow.  Our chosen speed scaling approximates the size dependence of gas drag \citep{whipple51-meteors}.  A range of ejection speeds for each grain size makes a more realistic looking coma \citep[e.g.,][]{combi94, jones08}.  Therefore, we pick velocities from a Gaussian distribution of width $\sigma$, but impose minimum ejection speeds by requiring $s_1>0$ to 10~\mps{} to avoid negative velocities and excessively narrow tails \citep[cf.][]{jones08}.  We assume a power-law differential size distribution: $dn/da\propto a^{N}$ at the nucleus.  The model runs adopt $N=-1$ to ensure a logarithmically uniform statistical sampling from $a_\mathrm{min}$ to $a_\mathrm{max}$.  Particles in a simulated observation can be weighted or removed to simulate different active areas, cone opening angles ($w$), and size distributions ($N\neq-1$).

  Simulations are scaled to a specific mass loss rate or total ejected mass, projected onto the sky, and calibrated to flux density units.  We absolutely calibrate the simulation assuming grains have a geometric albedo of 0.04 at 0.55~\micron{} (0.043\% in the $r$-band based on our measured color) and scatter light following the Schleicher-Marcus phase function\footnote{\url{https://asteroid.lowell.edu/comet/dustphase/}} \citep{schleicher98,marcus07b}.  The use of this phase function based on coma observations is an approximation since individual grains do not scatter light as the grain ensemble would.\footnote{The Schleicher-Marcus phase function is within 4\% of 0.036~mag~\inv{deg} over the phase angles considered in this paper ($<25\degr$).}

  Our general philosophy for the best-fit process is to start with only a few variable parameters, increasing the degrees of freedom as needed to approach an acceptable fit.  Lightcurve fitting may use a least-squares approach, but simulated images are qualitatively compared to the telescopic images to determine the degree of agreement, generally using 2D photometric contours.

  \section{Results} \label{sec:results}
  \subsection{Lightcurve}\label{sec:phot}
\begin{figure*}[h]
    \centering
    \includegraphics[width=\textwidth]{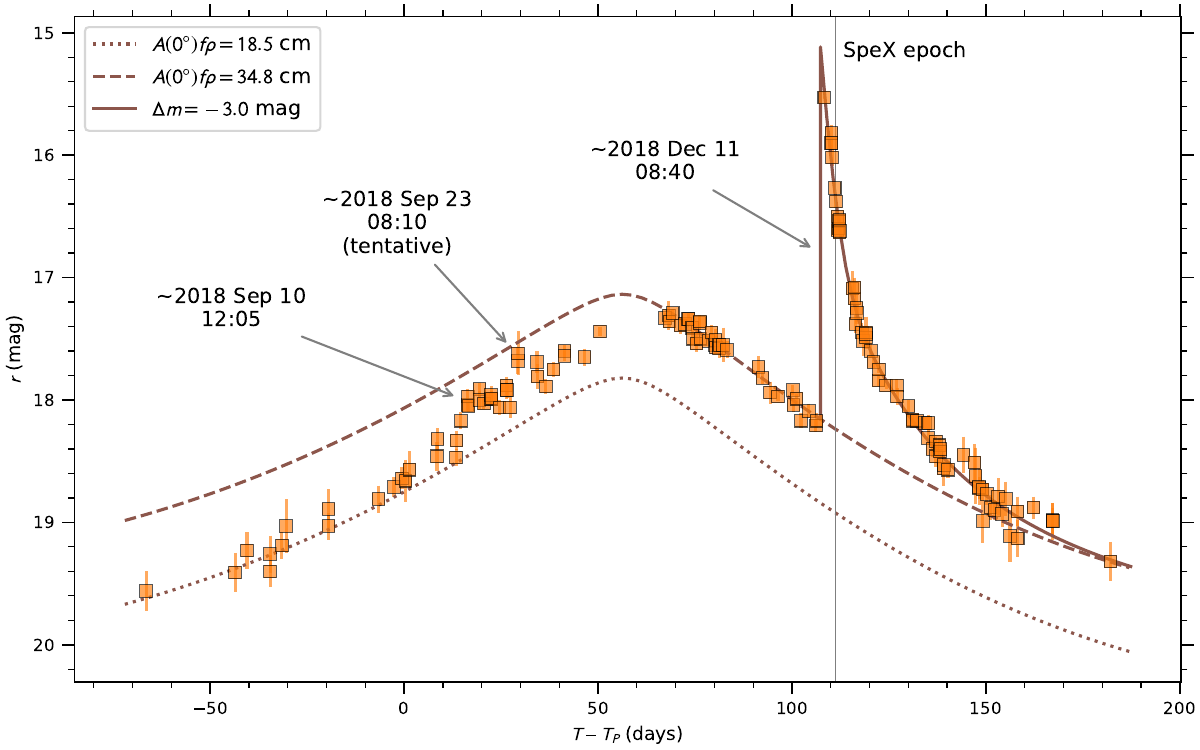}
    \caption{Lightcurve of comet 243P/NEAT versus time from perihelion ($T-T_P$) based on all photometry within a $10^4$~km radius aperture.  All data have been offset to the $r$-band using the measured colors of the coma (Section~\ref{sec:color}).  Two lines of constant \afrho[0\degr]{} are shown to emphasize the change in brightness as the comet moved through opposition at $T-T_P=57$~days (dotted and dashed lines).  Two outbursts are identified and labeled with their estimated dates (UTC), and a third is labeled as a tentative event.  A thin vertical line marks the time of our near-infrared spectrum with the SpeX instrument.  The best-fit lightcurve for the larger outburst is also shown (solid line).}
    \label{fig:lightcurve}
  \end{figure*}

\begin{figure*}[h]
    \centering
    \includegraphics[width=\textwidth]{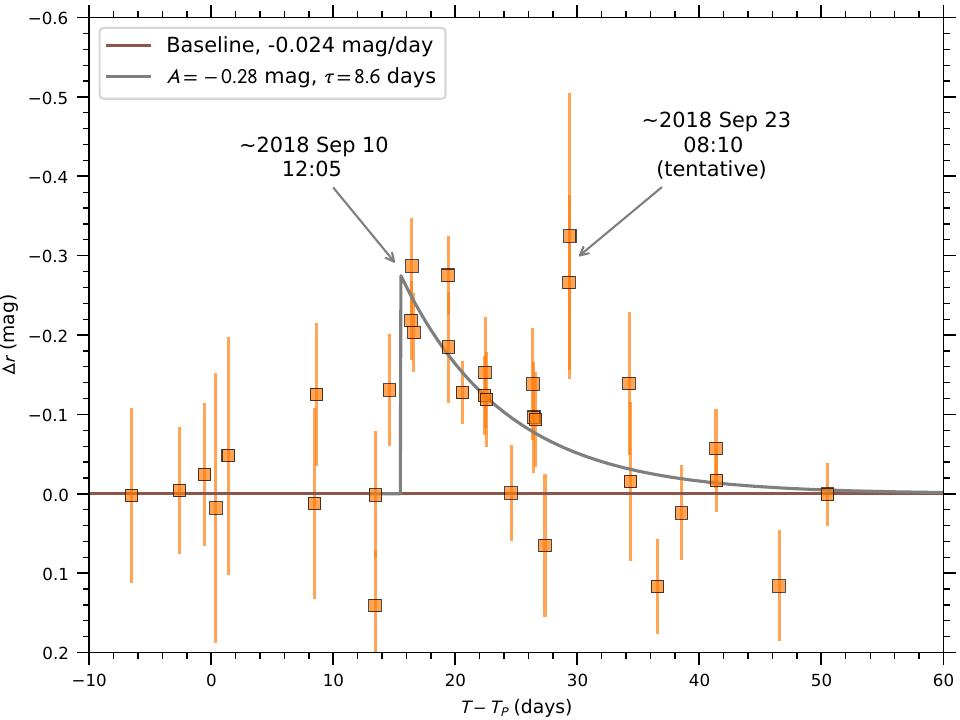}
    \caption{Photometry of comet 243P/NEAT showing a small outburst near 15 days after perihelion.  A slope of -0.024~mag~\inv{day} has been removed from the data.  A best-fit exponential curve with a fixed peak time of 2018 September 10 at 12:05 UTC, an amplitude of $-0.28\pm0.04$~mag and a $8.6\pm2.1$ day timescale is shown.  A second event may have occurred near $T-T_P=28$~days.}
    \label{fig:small}
  \end{figure*}

  Photometry of the comet is presented in Table~\ref{tab:obs} and plotted in Fig.~\ref{fig:lightcurve}.  Three features in the lightcurve are apparent: (1) the large outburst discovered by \citet{heinze18-cbet4587} at $T-T_P=107.35$~days, (2) small discontinuities at $T-T_P=15.50$ and 28.34~days, and (3) a substantial asymmetry in the lightcurve about opposition, which occurred at $T-T_P=57$~days (2018 October 22).  We identify the first small discontinuity as an outburst, verified by inspection of the individual images.  The strength was at least $-0.3$~mag in our 10,000-km photometric aperture, with an onset time of 2018 September 09 00:50 UTC and full uncertainty (i.e., 100\% confidence interval) of $\pm0.58$~days.  This event is followed by a tentative outburst between $T-T_P=27.35$ and 29.33~days (2018 September 23 08:19 UTC, $\pm0.99$~days), based on two photometric outliers (3.0$\sigma$ average deviation).  These smaller events are shown in detail in Fig.~\ref{fig:small}, where we plot the lightcurve after removing a slope of $-0.024$~mag~\inv{day}, derived from the photometry between $T-T_P=$0--14 and 41--51~days.

  For the large 2018 December outburst, the nominal 10,000-km aperture (7\farcs4) is smaller than the observed distribution at the time of the outburst discovery.  A 16\arcsec{} radius aperture encompasses the full outburst in the ATLAS discovery data and yields $o=15.31\pm0.02$~mag (corresponding to $r=15.44$~mag).  The outburst occurred between two ATLAS observations on December 10 and 12.  We use the mean date, $T_0=$2018 December 11 08:40 UTC, as a reference point and nominal time of outburst with a full uncertainty of $\pm0.98$~days.

  For context with other comets, we convert the magnitudes into the \afr{} quantity of \citet{ahearn84-bowell} and add the results to Table~\ref{tab:obs}.  The \afr{} quantity is the product of dust grain albedo $A$ (equal to 4 times the geometric albedo), effective filling factor of the dust $f$, and the photometric aperture radius $\rho$:
  \begin{equation}
    Af\rho = \frac{4\Delta^2}{\rho}\frac{F_\nu}{S_\nu(\rh)},
    \label{eq:afrho}
  \end{equation}
  where $\Delta$ is the observer-target distance, $F_\nu$ is the spectral flux density of the comet within the photometric aperture, and $S_\nu(\rh)$ is the spectral flux density of sunlight at the heliocentric distance of the comet.  The \afr{} value is independent of aperture size for a coma with surface brightness $\propto\rho^{-1}$.  It carries the units of $\rho$, which is expressed as a linear size at the distance of the comet.  Because $A$ has a phase angle dependence, \afr{} is commonly corrected to a phase angle of 0\degr{} and labeled \afrho[0\degr].  In the absence of gas contamination, and given assumptions on expansion speed and dust grain properties (density, size distribution), \afr{} can be converted into mass-loss rate \citep{fink12}.  Quiescent activity \afrho{} values are near 9~cm at the start of the data set, peaking at 27~cm near opposition with the Sun, and declining to 16~cm at the end of the data set.

  \subsection{Color}\label{sec:color}
  \subsubsection{243P/NEAT}
  The comet's photometric colors are listed in Table~\ref{tab:color}.  The colors are calculated from nightly averaged photometry, except for ATLAS, which tends to lack same-night color data.  Instead, ATLAS $c$-band measurements were compared to a local linear fit to the $o$-band magnitudes.  The average coma colors, $g-r=0.57\pm0.01$~mag and $c-o=0.46\pm0.02$~mag, were used to compute color corrections for the final photometric calibration.

  \begin{deluxetable*}{lc *{8}{r@{$\,\pm\,$}l}}
    \tabletypesize{\scriptsize}
    \tablecaption{Coma Optical Color Index and Spectral Gradient \label{tab:color}}
    \tablehead{
      \colhead{Source} & \colhead{Date} & \multicolumn2c{$RC-BC$} & \multicolumn2c{$g-r$} & \multicolumn2c{$B-V$} & \multicolumn2c{$V-R$} & \multicolumn2c{$R-I$} & \multicolumn2c{$c-o$} & \multicolumn2c{$S_V$\tablenotemark{a}} \\
      & \colhead{(UT)} & \multicolumn2c{(mag)}   & \multicolumn2c{(mag)} & \multicolumn2c{(mag)} & \multicolumn2c{(mag)} & \multicolumn2c{(mag)} & \multicolumn2c{(mag)} & \multicolumn2c{(\% per 100 nm)}
    }
    \decimals
    \startdata
    ZTF                   & 2018 Sep 08 10:29 & \nocolor    & 0.44 & 0.11 & \nocolor    & \nocolor    & \nocolor    & \nocolor    &  3.2 & 9.8 \\
    ZTF                   & 2018 Sep 11 09:23 & \nocolor    & 0.51 & 0.08 & \nocolor    & \nocolor    & \nocolor    & \nocolor    &  8.2 & 7.2 \\
    ZTF                   & 2018 Sep 14 09:55 & \nocolor    & 0.67 & 0.09 & \nocolor    & \nocolor    & \nocolor    & \nocolor    & 19.0 & 7.9 \\
    ZTF                   & 2018 Sep 17 09:36 & \nocolor    & 0.55 & 0.09 & \nocolor    & \nocolor    & \nocolor    & \nocolor    & 11.0 & 7.9 \\
    ZTF                   & 2018 Sep 21 08:37 & \nocolor    & 0.63 & 0.10 & \nocolor    & \nocolor    & \nocolor    & \nocolor    & 16.3 & 9.1 \\
    ZTF                   & 2018 Sep 29 06:13 & \nocolor    & 0.70 & 0.13 & \nocolor    & \nocolor    & \nocolor    & \nocolor    & 21.3 & 12.3 \\
    ZTF                   & 2018 Oct 06 09:37 & \nocolor    & 0.53 & 0.06 & \nocolor    & \nocolor    & \nocolor    & \nocolor    &  9.8 & 5.9 \\
    ZTF                   & 2018 Nov 07 06:32 & \nocolor    & 0.58 & 0.06 & \nocolor    & \nocolor    & \nocolor    & \nocolor    & 13.0 & 5.9 \\
    ATLAS                 & 2018 Nov 08 11:22 & \nocolor    & \nocolor    & \nocolor    & \nocolor    & \nocolor    & 0.50 & 0.05 & 17.2 & 4.2 \\
    ZTF                   & 2018 Nov 10 06:35 & \nocolor    & 0.60 & 0.07 & \nocolor    & \nocolor    & \nocolor    & \nocolor    & 14.0 & 6.6 \\
    ATLAS                 & 2018 Nov 12 10:57 & \nocolor    & \nocolor    & \nocolor    & \nocolor    & \nocolor    & 0.45 & 0.04 & 13.0 & 3.8 \\
    ZTF                   & 2018 Nov 14 07:17 & \nocolor    & 0.62 & 0.11 & \nocolor    & \nocolor    & \nocolor    & \nocolor    & 15.6 & 10.5 \\
    ZTF                   & 2018 Nov 15 06:27 & \nocolor    & 0.61 & 0.09 & \nocolor    & \nocolor    & \nocolor    & \nocolor    & 14.8 & 8.7 \\
    ATLAS                 & 2018 Dec 06 08:36 & \nocolor    & \nocolor    & \nocolor    & \nocolor    & \nocolor    & 0.56 & 0.06 & 22.2 & 5.6 \\
    ATLAS                 & 2018 Dec 10 08:50 & \nocolor    & \nocolor    & \nocolor    & \nocolor    & \nocolor    & 0.49 & 0.06 & 16.7 & 5.4 \\
    LCOGT                 & 2018 Dec 15 01:01 & \nocolor    & 0.52 & 0.06 & \nocolor    & \nocolor    & \nocolor    & \nocolor    &  9.0 & 5.7 \\
    NEXT\tablenotemark{b} & 2018 Dec 15 17:58 & \nocolor    & 0.60 & 0.03 & \nocolor    & \nocolor    & \nocolor    & \nocolor    & 14.3 & 2.8 \\
    LCOGT                 & 2018 Dec 15 19:05 & \nocolor    & 0.54 & 0.05 & \nocolor    & \nocolor    & \nocolor    & \nocolor    & 10.3 & 4.3 \\
    TRAPPIST-N            & 2018 Dec 15 21:39 & \nocolor    & \nocolor    & 0.95 & 0.14 & 0.42 & 0.14 & 0.43 & 0.14 & \nocolor    & 11.2 & 7.5 \\
    LCOGT                 & 2018 Dec 16 01:05 & \nocolor    & 0.59 & 0.06 & \nocolor    & \nocolor    & \nocolor    & \nocolor    & 13.8 & 5.2 \\
    LDT                   & 2018 Dec 16 02:25 & 1.64 & 0.03 & \nocolor    & \nocolor    & \nocolor    & \nocolor    & \nocolor    & 12.8 & 2.6 \\
    LDT (ROI a)\tablenotemark{c} & 2018 Dec 16 02:25 & 1.66 & 0.03 & \nocolor    & \nocolor    & \nocolor    & \nocolor    & \nocolor    & 13.6 & 2.8 \\
    LDT (ROI b)\tablenotemark{c} & 2018 Dec 16 02:25 & 1.57 & 0.04 & \nocolor    & \nocolor    & \nocolor    & \nocolor    & \nocolor    & 10.3 & 3.7 \\
    LDT (ROI c)\tablenotemark{c} & 2018 Dec 16 02:25 & 1.56 & 0.04 & \nocolor    & \nocolor    & \nocolor    & \nocolor    & \nocolor    & 10.0 & 3.7 \\
    LDT (ROI d)\tablenotemark{c} & 2018 Dec 16 02:25 & 1.56 & 0.06 & \nocolor    & \nocolor    & \nocolor    & \nocolor    & \nocolor    & 10.0 & 5.5 \\
    LDT (ROI e)\tablenotemark{c} & 2018 Dec 16 02:25 & 1.57 & 0.03 & \nocolor    & \nocolor    & \nocolor    & \nocolor    & \nocolor    & 10.3 & 2.8 \\
    LCOGT                 & 2018 Dec 16 10:09 & \nocolor    & 0.57 & 0.08 & \nocolor    & \nocolor    & \nocolor    & \nocolor    & 12.4 & 7.3 \\
    ZTF                   & 2018 Dec 20 02:53 & \nocolor    & 0.50 & 0.09 & \nocolor    & \nocolor    & \nocolor    & \nocolor    & 7.3 & 8.7 \\
    LCOGT                 & 2018 Dec 20 19:09 & \nocolor    & 0.64 & 0.15 & \nocolor    & \nocolor    & \nocolor    & \nocolor    & 16.9 & 13.6 \\
    LCOGT                 & 2018 Dec 22 00:49 & \nocolor    & 0.62 & 0.12 & \nocolor    & \nocolor    & \nocolor    & \nocolor    & 15.9 & 11.5 \\
    LCOGT                 & 2018 Dec 23 01:25 & \nocolor    & 0.59 & 0.13 & \nocolor    & \nocolor    & \nocolor    & \nocolor    & 13.9 & 11.8 \\
    ZTF                   & 2018 Dec 23 03:33 & \nocolor    & 0.58 & 0.14 & \nocolor    & \nocolor    & \nocolor    & \nocolor    & 12.7 & 12.8 \\
    TRAPPIST-N            & 2018 Dec 23 18:48 & \nocolor    & \nocolor    & 0.91 & 0.14 & 0.41 & 0.14 & 0.67 & 0.14 & \nocolor    & 18.4 & 7.4 \\
    LCOGT                 & 2018 Dec 25 00:50 & \nocolor    & 0.56 & 0.07 & \nocolor    & \nocolor    & \nocolor    & \nocolor    & 11.5 & 6.2 \\
    TRAPPIST-N            & 2018 Dec 25 22:46 & \nocolor    & \nocolor    & 1.20 & 0.14 & 0.40 & 0.14 & 0.40 & 0.14 & \nocolor    & 14.8 &  7.4 \\
    LCOGT                 & 2018 Dec 26 10:47 & \nocolor    & 0.57 & 0.05 & \nocolor    & \nocolor    & \nocolor    & \nocolor    & 12.1 & 4.4 \\
    ARC\tablenotemark{b}  & 2018 Dec 31 01:20 & \nocolor    & 0.51 & 0.09 & \nocolor    & \nocolor    & \nocolor    & \nocolor    & 8.2 & 8.7 \\
    TRAPPIST-N            & 2019 Jan 01 23:21 & \nocolor    & \nocolor    & 0.60 & 0.14 & 0.62 & 0.14 & 0.21 & 0.14 & \nocolor    & 5.0 & 7.4 \\
    ATLAS                 & 2019 Jan 07 07:38 & \nocolor    & \nocolor    & \nocolor    & \nocolor    & \nocolor    & 0.36 & 0.05 &  6.1 & 5.0 \\
    ZTF                   & 2019 Jan 04 02:58 & \nocolor    & 0.59 & 0.06 & \nocolor    & \nocolor    & \nocolor    & \nocolor    & 13.6 & 5.5 \\
    ZTF                   & 2019 Jan 08 02:38 & \nocolor    & 0.46 & 0.09 & \nocolor    & \nocolor    & \nocolor    & \nocolor    & 5.0 & 8.5 \\
    ZTF                   & 2019 Jan 10 02:37 & \nocolor    & 0.71 & 0.10 & \nocolor    & \nocolor    & \nocolor    & \nocolor    & 21.5 & 9.1 \\
    ZTF                   & 2019 Jan 11 02:38 & \nocolor    & 0.63 & 0.12 & \nocolor    & \nocolor    & \nocolor    & \nocolor    & 16.1 & 10.7 \\
    ATLAS                 & 2019 Jan 11 06:49 & \nocolor    & \nocolor    & \nocolor    & \nocolor    & \nocolor    & 0.41 & 0.08 &  9.9 & 7.1 \\
    TRAPPIST-N            & 2019 Jan 24 19:17 & \nocolor    & \nocolor    & 0.88 & 0.14 & 0.73 & 0.14 & 0.46 & 0.14 & \nocolor    & 21.7 &  7.4 \\
    TRAPPIST-S            & 2019 Feb 05 01:10 & \nocolor    & \nocolor    & 0.96 & 0.14 & 0.50 & 0.14 & \nocolor    & \nocolor    & 11.5 &  9.2 \\
    \enddata
    \tablenotetext{a}{Spectral gradient centered at $V$-band (0.55~\micron).  TRAPPIST values are based on an average of all color pairs.}
    \tablenotetext{b}{Observed in the $V$ and $R$ filters, but calibrated to PanSTARRS~1 $g$ and $r$.}
    \tablenotetext{c}{LDT regions of interest (ROI) are based on boxes (a) through (e) in Fig.~\ref{fig:color-map}.}
  \end{deluxetable*}

  To make an effective comparison between the colors measured with different bandpasses, we convert each value to a spectral gradient in units of \%~per~100~nm, centered at a wavelength of 0.55~\micron, which we name $S_V$.  Comet coma reflectances are nearly linear over our wavelength range \citep{jewitt86}, and centering our spectral gradients on the reflectance at a common wavelength is straightforward.  The spectral gradient between two filters with effective wavelengths $\lambda_1$ and $\lambda_2$ is
  \begin{equation}
    S_{2,1} = \frac{10^{0.4 (C - C_\sun)} - 1}{10^{0.4 (C - C_\sun)} + 1}\frac{2}{\lambda_2 - \lambda_1},
    \label{eq:S}
  \end{equation}
  where $C$ is the comet color index ($m_1 - m_2$) and $C_\sun$ is the solar color index in magnitudes between $\lambda_1$ and $\lambda_2$ \citep{ahearn84-bowell}.  Our solar colors and bandpass effective wavelengths are from \citet{willmer18-sun}, except for the ATLAS filters for which we used the \texttt{synphot} and \texttt{sbpy} softwares and the telescopic throughputs \citep{tonry18-atlas,mommert19-sbpy,synphot2018ascl} to compute $m_{\sun,c}=-26.71$~mag, $m_{\sun,o}=-26.98$~mag.  To convert the spectral gradient to one effectively normalized in the $V$-band, we assume the reflectance is linear throughout the optical and use the formula:
  \begin{equation}
    S_V= \frac{S_{2,1}}{1 + S_{2,1} (\lambda_V - \lambda_{2,1})},
    \label{eq:S_V}
  \end{equation}
  where $\lambda_V=0.55$~\micron{}, and $\lambda_{2,1}$ is the average of $\lambda_1$ and $\lambda_2$.  Uncertainties on the color indices are propagated through Eqs.~\ref{eq:S} and \ref{eq:S_V}.

  The LDT narrow-band filters have limited gas contamination \citep{farnham00} and provide the best single observation estimate of the dust optical color, \SV{13\pm3}, measured between 0.45 to 0.71~\micron.  All other colors, averaged by observatory, agree within $1\sigma$, and the weighted average of all the color measurements is \SV{(13\pm1)}.  We conclude that there is no significant gas contamination in our broad-band photometry.  In addition, the pre- and post-outburst ($0<T-T_0<15$~day) averaged color showed no difference: 
  \SV{12\pm2} and \specgrad{13\pm2}, respectively.

\begin{figure*}[t]
    \includegraphics[width=\textwidth]{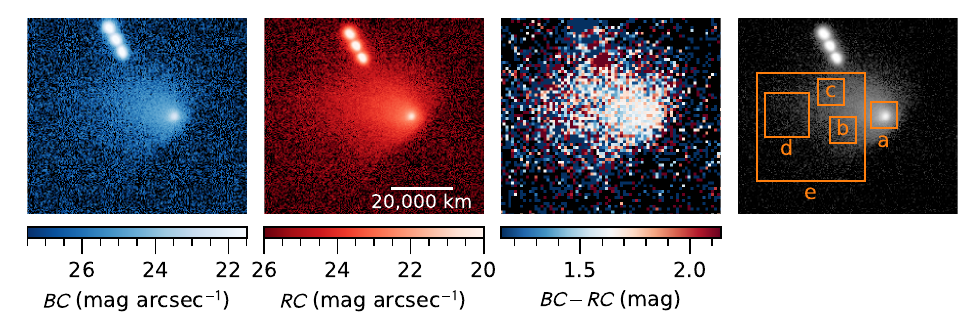}
    \caption{Images of comet 243P/NEAT and derived color, based on Lowell Discovery Telescope data taken 2018 December 16, $5\pm1$ days after a large outburst: (left) $BC$ filter image, (left middle) $RC$, (right middle) $BC-RC$ color index, and (right) regions of interest (a) through (e).  North is up, east to the left, and a scale bar is indicated.  The images are at the native pixel scale of the data (0\farcs24~pix$^{-1}$), but the color map has been 3$\times$3 binned.  Five square regions of interest are shown, centered at the following distances to the nucleus: (a) 0~km, (b), 15,000~km, (c) 20,000~km, (d) 33,000~km, and (e) 25,000~km.}
    \label{fig:color-map}
  \end{figure*}

  A color map produced from the LDT narrow-band data is presented in Fig.~\ref{fig:color-map}.  Five regions of interest are shown in the figure, and their mean colors are provided in Table~\ref{tab:color}.  The color difference between the near-nucleus and tail-ward ejecta, (a) versus (e), is only significant at the 2$\sigma$ level.  Altogether, no color gradient is seen, out to a distance of 40,000~km.

  \subsubsection{103P/Hartley 2}
\begin{figure*}
    \centering
    \includegraphics[width=\textwidth]{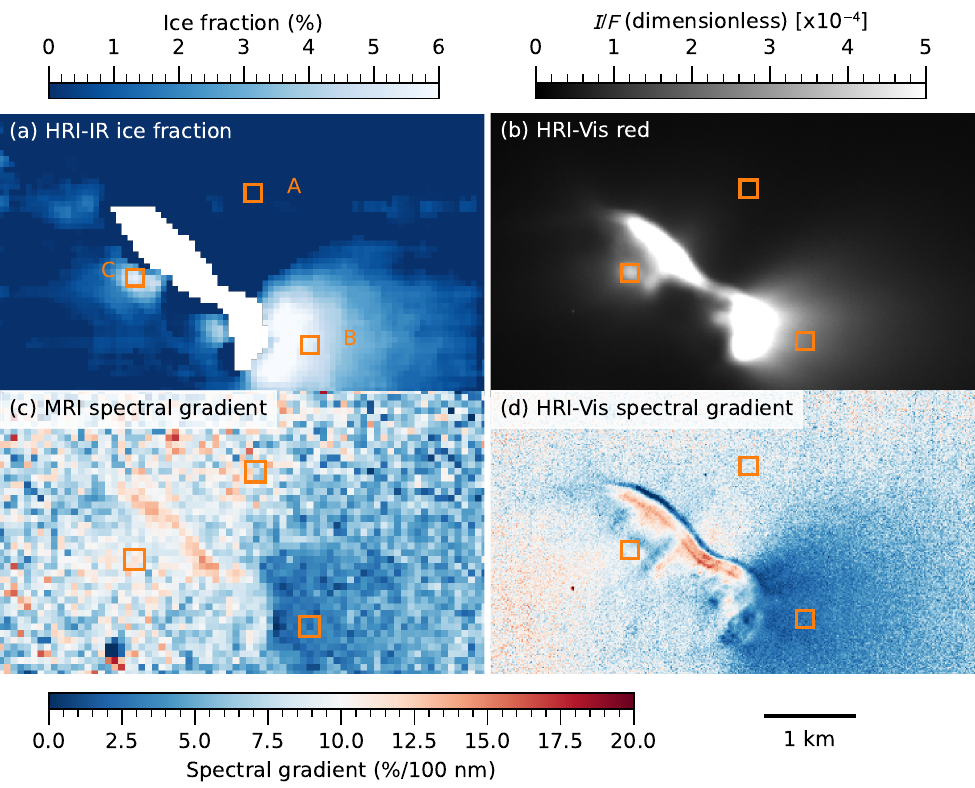}
    \caption{Inner-coma maps of comet 103P/Hartley 2 derived from Deep Impact data, oriented with the Sun to the right.  A linear scale bar is shown in the lower right.  (a) Water ice fractional abundance by area from \citet{protopapa14}.  The illuminated nucleus surface has been masked.  Boxes A and B from \citet{protopapa14} and an additional Box C (chosen by us) are shown in all panels.  (b) HRI-Vis red filter image for context, displayed as $I/F$ (ratio of the radiance from the comet to the incident radiance from the Sun).  (c) MRI spectral gradient map from the green continuum and red filters, with the color scale indicated in the figure.  (d) HRI-Vis spectral gradient from the blue and red filters, using the same color scale as the MRI data.}
    \label{fig:di-color}
  \end{figure*}

  Color maps generated from the Deep Impact observations of comet 103P/Hartley 2 and expressed as $S_V$ are presented in Fig.~\ref{fig:di-color}.  Also shown is the ice abundance map derived by \citet{protopapa14} based on IR spectral imaging.  Two regions of interest are highlighted in the figure, Boxes A and B of \citet{protopapa14}.  Box A is depleted in water ice grains ($<$0.1\% by area), and Box B is enriched in water ice (5\% by area).  We additionally investigated a third region, Box C, centered on a small but bright and icy outflow of material, labeled ``Jet 4'' ($J_4$) by \citet{protopapa14}.

  The mean spectral gradients for MRI and HRI-Vis are given in Table~\ref{tab:slopes-and-bands}.  Uncertainties are based on the scatter of the color data within each box.  \citet{klaasen08} quotes an absolute calibration uncertainty of 5 and 10\% for the HRI-Vis and MRI filters used here, respectively, which corresponds to spectral gradient uncertainties of 2.5 and \specgrad{6.9}.  Although the absolute slopes may carry large uncertainties, the relative change in slope between the boxes should be secure.  Regardless of the instrument used, Box B is the bluest ($S_V \sim 2$\% per 100 nm), and Boxes A and C have similar colors ($S_V \sim 9$\% per 100 nm).

  \begin{deluxetable*}{lcccccccc}
    \tablecaption{Continuum Slope, 2.0-\micron{} Water Ice Band Depth, Modeled Areal Fraction, and Grain Size \label{tab:slopes-and-bands}}
    \tablehead{
      \colhead{Comet} & \colhead{$\lambda_b$} & \colhead{$\lambda_r$} & \colhead{$S_V$} & \colhead{2-\micron{} band depth} & \colhead{Ice areal fraction} & \colhead{Ice grain radius} \\
      & \micron & \micron & (\specgrad{}) & (\%) & (\%) & \micron
    }
    \startdata
    243P/NEAT                   & 0.45 & 0.71 & $12.8\pm2.6$ \\
    &  0.9 &  1.4 & $4.4\pm0.3$ \\
    &  1.7 &  2.25 & $1.0\pm0.3$ & $<3$ &  & \\\hline
    C/2013 US$_{10}$ (Catalina) &  0.9 &  1.3 & $1.9\pm0.2$ \\
    &  1.7 &  2.25 & $-0.1\pm0.2$ & $11.0\pm0.9$ & $15\pm1$ & 0.6 \\\hline
    103P/Hartley 2 (Box A)      & 0.45 & 0.74 & $8.5\pm0.1$ \\
    & 0.53 & 0.74 & $8.2\pm0.8$ \\
    &  1.7 &  2.25 & $3.4\pm0.6$ & $<6$ & $<0.1$ & (0.5) \\\hline
    103P/Hartley 2 (Box B)      & 0.45 & 0.74 & $2.6\pm0.1$ \\
    & 0.53 & 0.74 & $3.2\pm0.1$ \\
    &  0.9 &  1.4 & $1.8\pm0.4$ \\
    &  1.7 &  2.25 & $1.2\pm0.1$ & $9.8\pm0.5$ & $5.3\pm0.2$ & 0.4 \\\hline
    103P/Hartley 2 (Box C)      & 0.45 & 0.74 & $8.5\pm0.1$ \\
    & 0.53 & 0.74 & $10.3\pm0.4$ \\
    &  0.9 &  1.4 & $6.6\pm0.7$ \\
    &  1.7 &  2.25 & $3.4\pm0.2$ & $6.3\pm0.4$ & 3.7 & (0.5) \\
    \enddata
    \tablecomments{$\lambda_b$, $\lambda_r$ are the blue and red filters (optical) or the wavelength range (infrared) used to determine the spectral slope.  $S_V$ is the spectral slope, normalized to 0.55~\micron.  Upper-limits are quoted for the 3$\sigma$ level.  The C/2013 US$_{10}$ quantities are measured from the \citet{protopapa18-catalina} spectrum at 5.8~au.  103P box labels refer to Fig.~\ref{fig:di-color}.  Model water ice grain parameters are from \citet{protopapa14,protopapa18-catalina}.  Values in parentheses are assumed.}
  \end{deluxetable*}

  \subsection{Morphology}

\begin{figure*}
    \centering
    \includegraphics[width=\textwidth]{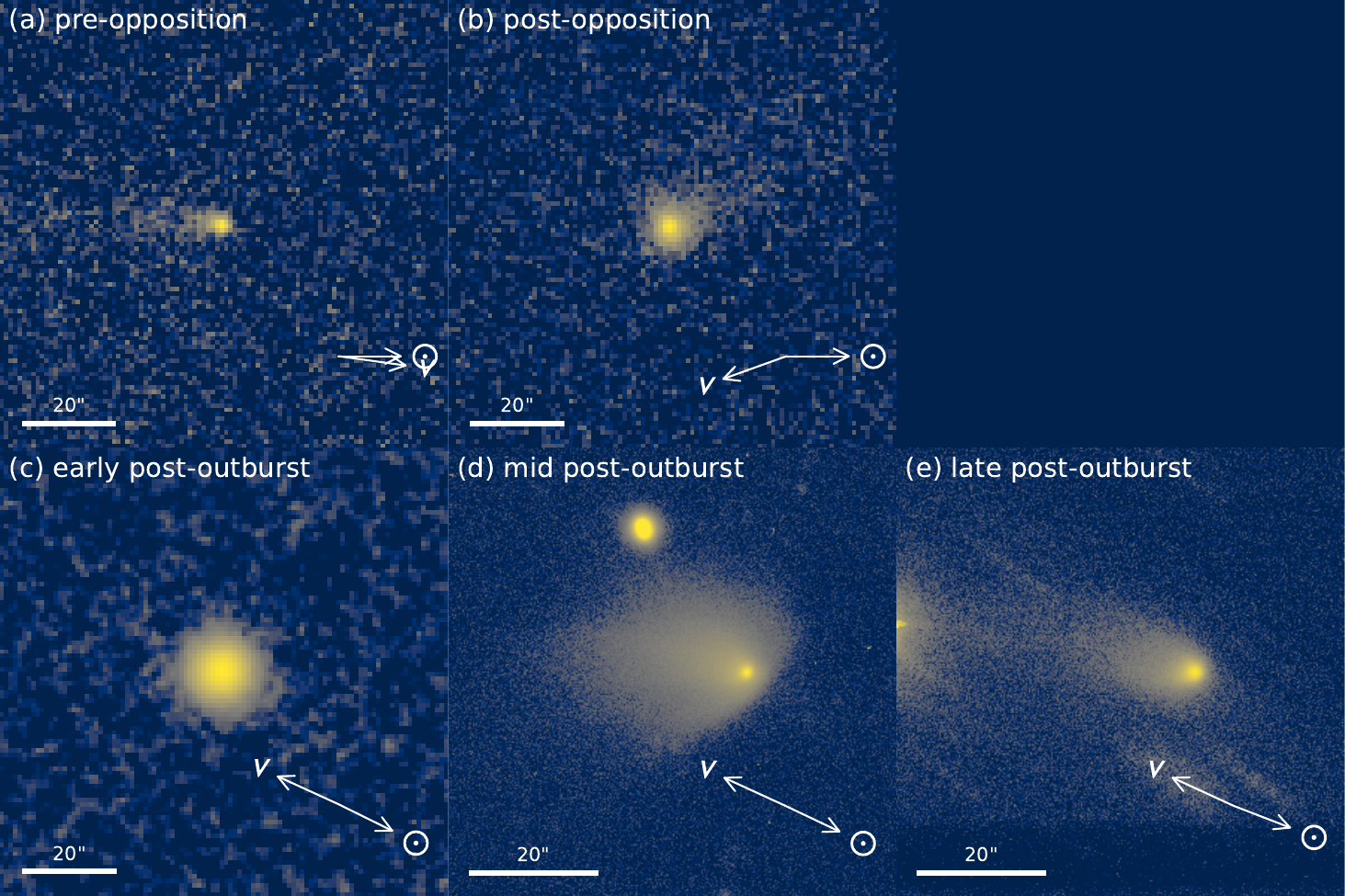}
    \caption{Select images of comet 243P/NEAT: (a) median ZTF pre-opposition $r$-band image; (b) median ZTF post-opposition $r$-band image; (c) mean ATLAS $o$-band image, 0.99~days post-outburst (resampled to 1\farcs0 pix$^{-1}$); (d) single LDT $r'$-band image, 4.73~days post-outburst; and, (e) mean LDT $r'$-band image, 31.73~days post-outburst.  The ZTF images are generated from multiple nights of data aligned with the projected comet-Sun vector on the x-axis (see text for details).  The ATLAS and LDT images are aligned with celestial north up and east to the left.  Color scales for all images linearly vary from 0 to the 3$\sigma$ background level, then logarithmically up to the peak pixel.  Projected Sun and velocity vectors and scale bars are shown (averages for the ZTF data).}
    \label{fig:morph}
  \end{figure*}

  Our pre-outburst images are based on survey data, and they have limited morphological information due to the short exposure times and relative faintness of the target.  During this period, the comet moved through opposition, which can affect its morphology and brightness, particularly the projection of the tail on the sky.  To investigate this effect, we median combine ZTF $r$-band images into pre- and post-opposition stacks: 15 images from 2018 July 02 to August 27 (avoiding the September outbursts), and 14 images from 2018 November 01 to November 25 (Fig.~\ref{fig:morph}).  The images were rotated to align the projected comet-sun vector before being combined.  A short tail is evident in both stacks.  Over opposition, the tail rotated from the anti-sunward direction to $\sim40$\degr{} from the sunward vector.  We also aligned and combined the images in the rest frame of the comet's projected velocity vector, and the morphology was essentially the same.

  Figure~\ref{fig:morph} shows the morphological evolution of the December 2018 outburst from the discovery image to 32 days after the outburst.  The discovery image shows a nearly azimuthally symmetric compact source, with a FWHM larger than nearby point sources of similar brightness, 7\farcs4 versus 6\farcs3, the latter measured on a mean image aligned in the celestial reference frame.  A crude estimate of the ejecta motion in the plane of the sky may be made assuming the ejecta's surface brightness is Gaussian-like: $(7.4^2 - 6.3^2)^{1/2} \sim 4\arcsec$ (5000 km).  Two images from the LDT show the post-outburst morphological evolution from a broad extended source to one resembling a cometary dust tail.  We identify the broad component as the outburst ejecta.

  \subsection{Spectroscopy}\label{sec:results:spectroscopy}

\begin{figure*}
    \centering
    \includegraphics[width=\textwidth]{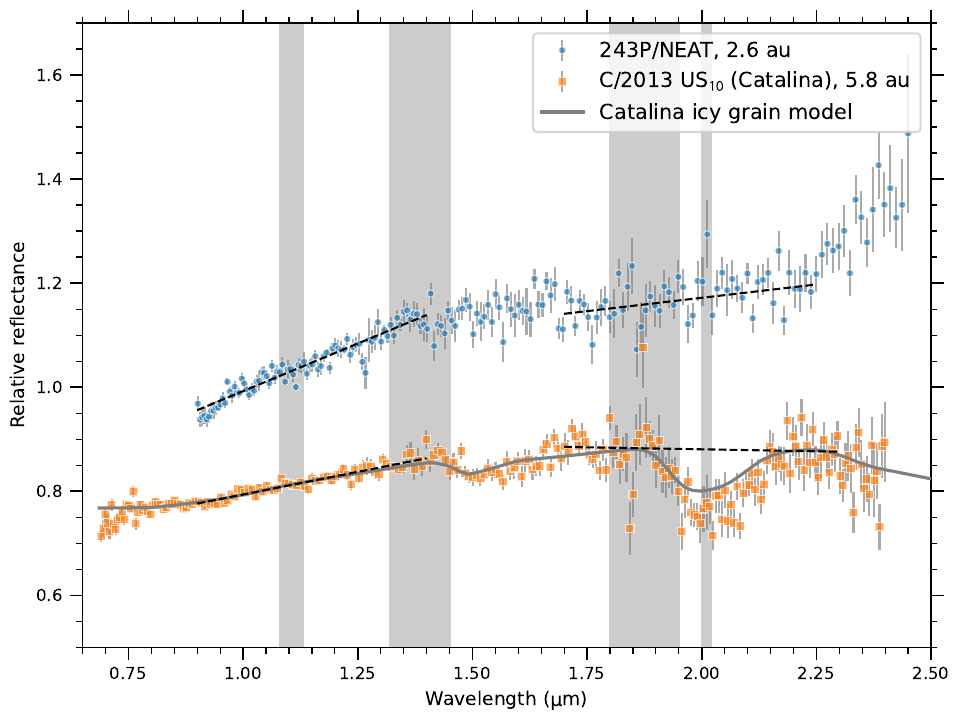}
    \caption{IRTF/SpeX spectra of comet 243P/NEAT at $\rh=2.6$~au, taken 2018 December 15, $4\pm1$ days after a large outburst (circles).  Also shown is an IRTF/SpeX spectrum of comet C/2013 US$_{10}$ (Catalina) at 5.8~au (squares).  Both spectra are normalized to 1.0 at 1.0~\micron, but the Catalina spectrum has been offset by $-0.2$ for clarity.  The Catalina spectrum is from \citet{protopapa18-catalina}, and serves as an example for the 1.5- and 2.0-\micron{} water ice absorption features seen in some comets, but are absent from the 243P spectrum.  The icy grain model for comet Catalina from \citet{protopapa18-catalina} is also shown (solid line).  Shaded regions indicate wavelengths that may be affected by strong telluric absorption.  The spectrum of comet 243P is available as Data Behind the Figure.}
    \label{fig:irtf}
  \end{figure*}

  The near-infrared reflectance spectrum of 243P taken 4 days after the outburst is shown in Figure~\ref{fig:irtf}. The spectrum has been binned to a spectral resolution of $\lambda/\Delta\lambda=190$, i.e., 2.5 times the spectral resolving power of $\sim$75.  The spectrum is red, and consists of three near-linear segments: $\lambda<1.4$, 1.4--2.25, and $>2.25$~\micron.  The latter segment indicates an increased reddening at the long wavelengths.  This feature persisted in all of our reduction experiments varying standard stars, extraction aperture, background subtraction, and atmospheric corrections.  It is not due to thermal emission from the dust or nucleus, given the heliocentric distance \citep[compare with comet 46P observations at 1.1~au;][]{protopapa21,kareta23}.  Although we suspect it is a calibration artifact, we cannot prove it, and leave it in the spectrum, but discuss it no further.

  For comparison to the 243P spectrum, we also show in Fig.~\ref{fig:irtf} a spectrum of comet C/2013 US$_{10}$ (Catalina) taken with the same instrument from \citet{protopapa18-catalina}.  The spectrum of Catalina clearly shows 1.5- and 2.0-\micron{} water-ice absorption features.  These features are absent from the 243P spectrum.

\begin{figure*}
    \centering
    \includegraphics[width=0.6\textwidth]{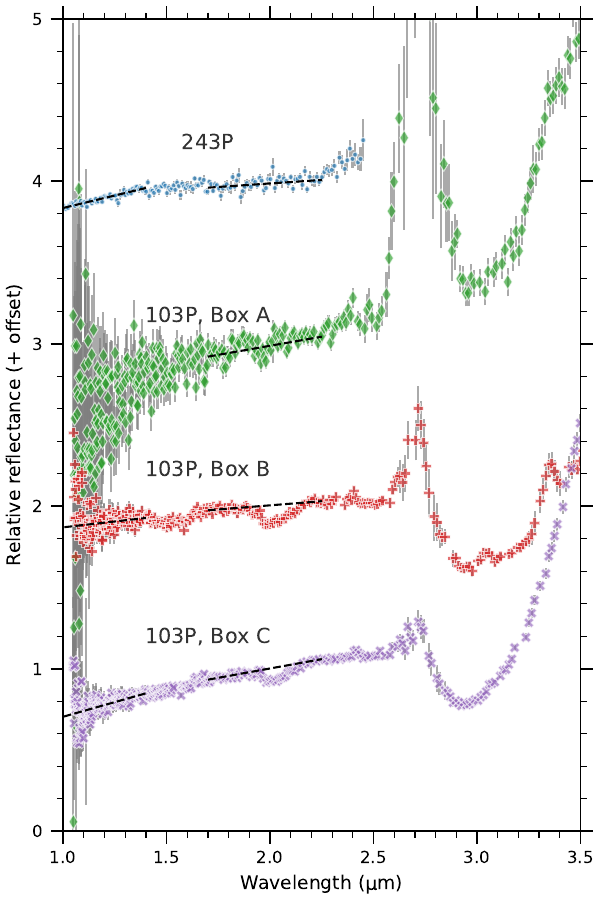}
    \caption{Deep Impact/HRI-IR spectra of comet 103P/Hartley 2 corresponding to Boxes A, B, and C in Fig.~\ref{fig:di-color}.  Also shown is the IRTF/SpeX spectrum of comet 243P/NEAT from Fig.~\ref{fig:irtf}.  Spectra are offset by a constant value for clarity.  Dashed lines indicate the best-fit linear slope to the continuum at 1.45--1.8 and 2.2--2.4~\micron.  Spectra are normalized to 1.0 at 2.0~\micron{}, based on the linear fit.  Identified water ice absorption features: 243P, none; 103P Box A, none; Box B, 1.5, 2.0, and 3~\micron{}; and Box C, 2.0 and 3~\micron.}
    \label{fig:di-spectra}
  \end{figure*}

  Near-infrared spectra of the inner coma of comet 103P show both icy and ice-free spectra \citep{protopapa14}.  Spectra extracted from the Deep Impact HRI-IR spectral map are shown in Fig.~\ref{fig:di-spectra} for each of the three regions of interest (Boxes A, B, and C) from Section~\ref{sec:color}.  The spectra of all boxes are dominated by scattered light ($<2.5$~\micron) and thermal emission ($>3.2$~\micron).  Water-ice absorption features are seen in the Box B and Box C spectra at 2.0 and 3~\micron, but the 1.5-\micron{} water ice band is only evident in the Box B spectrum.  Fluorescence emission from the $\nu_2$ rotational band of water gas is seen in all three boxes at 2.7~\micron{}, and emission from a variety of ``organic'' molecules is seen at 3.4~\micron{} in the Box A and B spectra \citep{protopapa14}.

  Outside of the absorption and emission features, the scattered light continuum is red and nearly linear in each of the five comet spectra.  Linear fits and approximate uncertainties were used to derive spectral slopes over two wavelength ranges using a modified Levenberg-Marquardt algorithm \citep[\texttt{leastsq} from the scipy Python package;][]{virtanen20-scipy}.  The first range, 0.9 to 1.4~\micron, was chosen based on the shape of the comet 243P spectrum, which appears to have a break in the slope at 1.4~\micron.  The second range, 1.7 to 2.25~\micron, was chosen to define a reference continuum for calculating the depth of the 2.0-\micron{} water-ice band, while avoiding the apparent increase in reflectivity at 2.25--2.40~\micron.  The wavelengths between 1.88 and 2.2~\micron{} were excluded from the fits.  The signal-to-noise ratio of the 103P Box A spectrum was too low for a slope measurement in the first segment.  The best-fit spectral slopes are reported in Table~\ref{tab:slopes-and-bands}.

  We calculated the depth of the 2.0-\micron{} water-ice band by comparison of the average reflectance at 1.95--2.05~\micron{} to the best-fit continuum at 1.7--2.3~\micron{}.  The results are reported in Table~\ref{tab:slopes-and-bands}.  Consistent with our visual inspection, the 2.0-\micron{} band is not detected in the 243P data and only an upper-limit is reported.  Also included in the table are the icy grain modeling results for comets Catalina and 103P.  The results for Box C are based on the modeled ice and dust maps from \citet{protopapa14} (Fig.~10 of their paper), which assumed a grain radius of 0.5~\micron.

  \section{Analysis}\label{sec:analysis}
  \subsection{Evidence for Water Ice}

  The spectrum of comet 243P lacks any evidence for the 1.5- and 2.0-\micron{} water-ice absorption features occasionally observed in comets \citep[e.g.,][]{kawakita04-ice}.  The 3$\sigma$ upper-limit 2.0-\micron{} band depth, 3\%, is meaningful when compared to the clear detections of ice in the comae of comets Catalina and 103P with band depths from 6 to 11\%.  An upper-limit ice fraction in 243P may be calculated by linearly extrapolating the Catalina and 103P ice fraction to band-depth ratios.  Given these two examples, the areal ice fraction in the ejecta of comet 243P is $\lesssim2-5$\%.  The estimates assume that the ice and dust properties of 243P are comparable to those of Catalina and 103P, which is sufficient as a first-order approximation.

  As an alternative approach, we inspect the coma continuum color for indirect evidence of water ice.  Bluing of the light scattered by comets has long been considered a possible signature of water ice \citep[e.g.,][]{fernandez07,jewitt15-color}.  However, a relationship between coma optical color and the presence of water ice is not well established.  In contrast, surfaces are better studied and show a clear bluing and brightening with increasing water ice content \citep[e.g.,][]{sunshine06,li13-hartley2,fornasier15-nucleus}.  Yet, our analysis of comet 103P lacks a singular relationship between the presence of ice and the scattered light continuum color.  At first, comet 103P's icy Box B is significantly bluer than the ice-free Box A, which agrees with the expectation that the presence of ice blues coma continuum color.  \citet{laforgia17-oh} arrive at a similar conclusion based on the comparison of the sunward and anti-sunward direction of the coma of comet 103P.  But, Box C breaks this relationship.  Icy Box C has nearly the same red color as ice-free Box A.  The difference must lie in the scattering properties of the dust grains.  Therefore, it appears that the dust composition (grain size, material, and/or porosity) varies with active region on this comet.

  Returning to 243P, LDT narrow-band colors taken $5\pm1$~days after the outburst showed no evidence for color variations out to 40,000~km, and were consistent with all other color measurements in our data set.  Adding the lack of water ice spectral features in the near-infrared spectrum, we conclude there is no evidence for water ice or any compositional gradient in the outburst ejecta, at least at 5~days after the outburst or later.

  \subsection{Quiescent Activity}\label{sec:quiesence}

  The peak of our quiescent lightcurve occurs near 68~days after perihelion.  This peak is related to the changing phase (Sun-comet-observer) angle and geocentric distance; the comet passed through opposition at $T-T_P=57$~days at a minimum phase angle of 4\degr.  To illustrate, we plot photometric models for an isotropic dust coma with a constant dust production rate in Fig.~\ref{fig:lightcurve} based on the \afr{} formalism (Eq.~\ref{eq:afrho}).  The models predict a peak brightness at the minimum phase angle, and a symmetric lightcurve about the opposition.  However, the post-opposition data are consistently brighter than the pre-opposition data, contrary to the isotropic model prediction.  In terms of \afr, the comet varies from \afrho[0\degr]=$18.5\pm1.0$~cm at $T-T_P=-45$ to $-30$~days, to $\afrho[0\degr]=34.8\pm0.3$~cm at +90 to +110~days.

  Rather than being a true increase in activity, the post-opposition brightening may be caused by slow-moving grains that build up near the nucleus, or projection effects due to the changing observation geometry of Earth-based observers.  The change in morphology in the ZTF images (Fig.~\ref{fig:morph}a and b) provides some evidence for this latter hypothesis.  We further tested it using our dust dynamical model (Section~\ref{sec:dynamical-model}).

  We set out to reproduce: (1) a post-opposition intrinsic brightening of --0.7~mag (Fig.~\ref{fig:lightcurve}); (2) a nearly linear, pre-opposition tail (Fig.~\ref{fig:morph}a); and, (3) a potentially broader tail pointed towards position angle $+40\degr$ from the Sun (Fig.~\ref{fig:morph}b).  To that end, isotropic models (global and uniform dust ejection) were generated for select ejection speeds, with dust production begining near aphelion (November 2014).

  A series of low-resolution ($10^7$ particle) models were converted into synthetic photometry and compared to the ZTF lightcurve, and higher-resolution ($7.5\times10^7$ particle) models were generated at 10 and 70 days post-perihelion, imaged, and compared to the ZTF image stacks.  Initial simulations producing model lightcurves and images were qualitatively compared to the data to determine the parameter sets to explore in detail.  Ejection speeds tested in detail were $s_1$=38, 50, and 100~\mps.  Particles were scaled to simulate various ejection anisotropies, dust production variations proportional to $r_h^k$ (for $k=-1$, --2, --3, and --6), and differential grain size distributions proportional to $a^N$ (for $N=-3.3$, --3.5, and --3.7; cf. \citealt{agarwal24-comets3}).  Two large grain size cutoffs were tested: 1~mm and 1~cm.

  To explore the effects of ejection anisotropies, we generated lightcurves and images for isotropic ejection, sunward directed activity, and for a set of 26 discrete areas nearly uniformly distributed in RA and Dec ($\approx$34\degr{} spacing).  Each directed source was tested with cone opening angles $w=30$, 45, 90, and 120\degr{}.  Out of the 26 active areas considered, we rejected those that were on the night side of the nucleus (Sun-zenith angles $>90\degr$).  Although night side dust production has been observed at comets due to nucleus rotation and surface thermal inertia \citep{farnham07, ahearn11, fougere16-sources}, here we would require substantial activity over a 100-day period from an unilluminated source.  Two of the remaining areas produced good morphologies, having an (RA, Dec) of ($210$\degr, $-10$\degr), and (280\degr, $-20$\degr).  Expressing these coordinates as ($I$, $\nu$), where $I$ is the angle from the orbital angular momentum vector and $\nu$ is the angle from perihelion in the orbital plane (i.e., true anomaly), they become ($90$\degr, $-20$\degr) and ($90$\degr, $-90$\degr).  The lightcurve for (280\degr, $-20$\degr) was nearly constant in apparent magnitude after opposition, and not a viable solution.  The ($210$\degr, $-10$\degr) model source may be interpreted as a high-latitude active area on a nucleus with a large obliquity.  It has a post-perihelion Sun-zenith angle ranging from 4 to 35\degr{}, i.e., consistently illuminated and within 35\degr{} of the sub-solar point of a spherical nucleus.

  We generated 648 model lightcurves based on combinations of the above parameter sets.  Using a least-squares fitting technique, each model was linearly scaled to match the quiescent lightcurve.  An examination of the best 100 models ($1.6<\chi^2_\nu<2.1$, 34 degrees of freedom) showed a preference for heliocentric distance slopes $k>=-3$: out of the 100 models, $k=-1$, --2, --3, or --6 was selected 29, 27, 27, and 17 times, respectively.  There was a slight preference for $a_\mathrm{max}=1$~mm (57/100 models).  Focusing on the $k=-2$ cases, there is a slight preference for differential grain size distribution slopes of $N=-3.3$ (16 out of 27), with the remaining models favoring $N=-3.5$.  The production distributions are all centered sunward or towards ($210$\degr, $-10$\degr) without preference, but low ejection speeds were preferred (15/27 with $s_1=38$~\mps, 8/27 with 50~\mps, and 4/27 with 100~\mps).  The pre-opposition brightening is well-fit by some models, but the post-opposition model lightcurves tend to be 0.05 to 0.10~mag too faint compared to the observations.  Given that our source location spacing is $\approx$34\degr{}, a small $\sim10\degr$ change in the production direction or some additional production asymmetry is needed to fully fit the data.

  We next simulated pre- and post-opposition images based on the best twelve $k=-2$ cases, and compared them by visual inspection of isophotal contours to the ZTF pre- and post-opposition image stacks.  No single model accounts for the imaging data, but there are some close matches.  Overall, the ($210$\degr, $-10$\degr) model source performed better than the sunward emission models in the morphology tests.  Specifically, the ($210$\degr, $-10$\degr) models better agreed with both the orientation and width of the pre-opposition tail, and the post-opposition coma at position angles around $-90$\degr{} from the Sun.  An examination of the $k=-1$, --3, and --6 cases did not reveal significant improvements, a sign that the small scale ($\sim10$\arcsec) morphology is dominated by recently produced dust.

\begin{figure*}
    \centering
    \includegraphics[width=\textwidth]{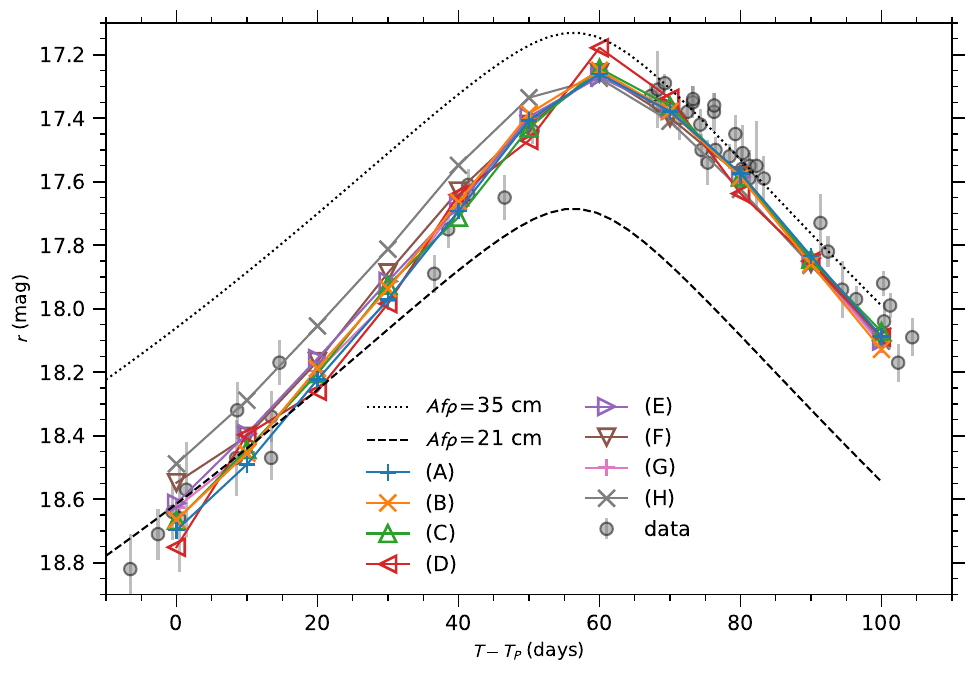}
    \caption{Photometry (circles) and model lightcurves (other symbols) of comet 243P/NEAT versus time from perihelion.  Labels correspond to the quiescent models in Table~\ref{tab:model}.  Photometry of the small September outbursts have been removed.  Most models have generally good agreement with the photometry, although they are all approximately 0.05--0.1~mag too faint after opposition ($T-T_P=57$~days).  In addition, the isotropic model (H) is too bright for the pre-opposition epochs.  The two lines of constant \afr{} (dotted and dashed lines) are also shown to emphasize the lightcurve asymmetry about opposition.}
    \label{fig:quiescent-lightcurve}
  \end{figure*}

\begin{figure*}
    \centering
    \includegraphics[width=\textwidth]{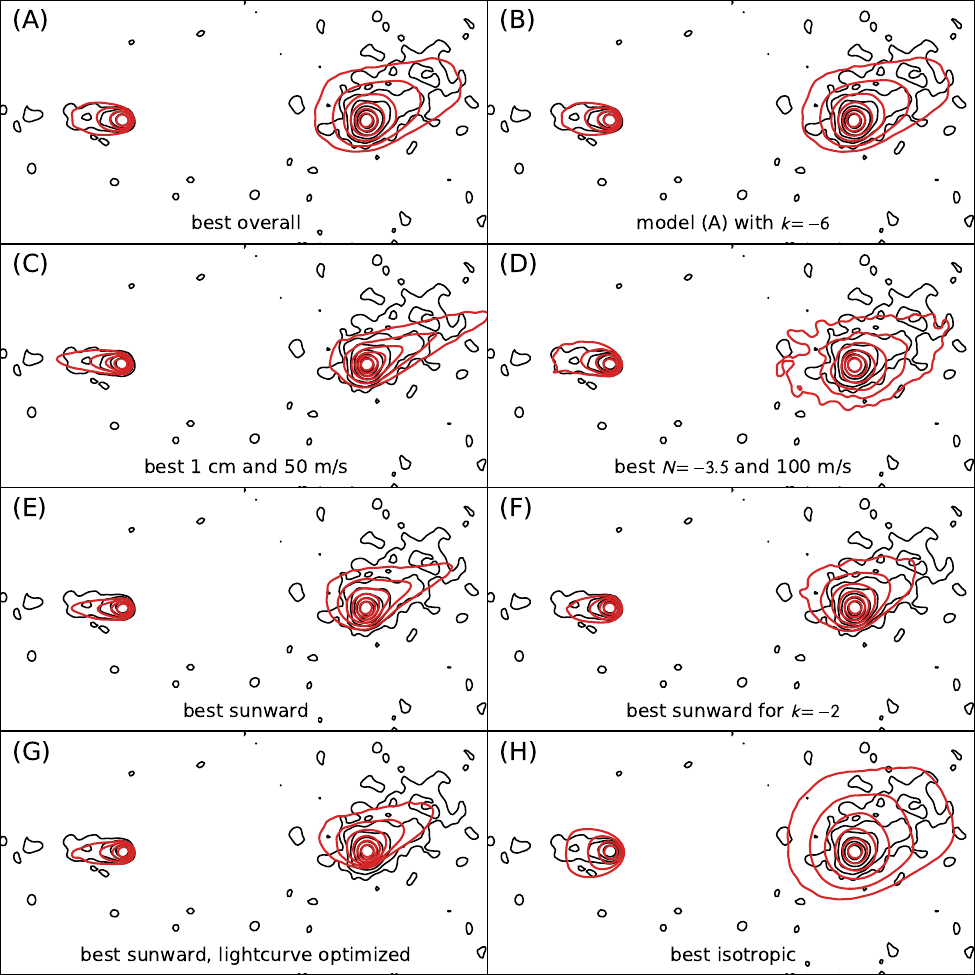}
    \caption{Morphology of the ZTF observations (black contours) pre-opposition (sub-panel, left) and post-opposition (sub-panel, right) compared to our example models (red contours).  Contour levels are with respect to the peak pixel of the comet and spaced at factors of two intervals from $2^{-1}$ to $2^{-4}$ (pre-opposition) or $2^{-6}$ (post-opposition).  The images were first smoothed with a $\sigma=1.5\arcsec$ Gaussian.  Image orientation is the same as in Fig.~\ref{fig:morph} (Sun to the right), and a 100\arcsec$\times$100\arcsec{} region ($\sim120,000\times120,000$~km) is shown around each image.  Panel labels correspond to the quiescent models in Table~\ref{tab:model}.}
    \label{fig:ztf}
  \end{figure*}

  Ten models (A--H) have been selected to demonstrate the effects of ejection anisotropy, speed, and grain size distribution.  In Figs.~\ref{fig:quiescent-lightcurve} and \ref{fig:ztf}, we compare the models to the comet's lightcurve and morphologies.  The model parameters are given in Table~\ref{tab:model}.

  \begin{deluxetable*}{cccccCCcc}
    \tablecaption{Dynamical Model Parameter Sets \label{tab:model}}
    \tablehead{
      \colhead{Model} & \colhead{$s_1$} & \colhead{$\sigma$} & \colhead{$\hat{v}$} & \colhead{$w$} & \colhead{$k$} & \colhead{$N$} & \colhead{$a_{\mathrm{min}}$} & \colhead{$a_{\mathrm{max}}$} \\
      & \colhead{(m \inv{s})} & \colhead{(m \inv{s})} & & \colhead{(\degr)} & & & \colhead{(\micron{})} & \colhead{(mm)}
    }
    \startdata
    \multicolumn{7}{l}{Quiescent coma}\\\hline
    A & 38 & 5 & 210\degr,$-10$\degr & 90 & -2 & -3.3 & 0.1 & 1 \\
    B & 38 & 5 & 210\degr,$-10$\degr & 90 & -6 & -3.3 & 0.1 & 1 \\
    C & 50 & 10 & 210\degr,$-10$\degr & 45 & -6 & -3.3 & 0.1 & 10 \\
    D & 100 & 20 & 210\degr,$-10$\degr & 30 & -2 & -3.5 & 0.1 & 1 \\
    E & 38 & 5 & sunward & 45 & -6 & -3.3 & 0.1 & 10 \\
    F & 50 & 10 & sunward & 45 & -2 & -3.3 & 0.1 & 1 \\
    G & 38 & 5 & sunward & 45 & -6 & -3.3 & 0.1 & 1 \\
    H & 38 & 5 & isotropic & \nodata & -1 & -3.3 & 0.1 & 1 \\\hline
    \multicolumn{7}{l}{Outburst ejecta}\\\hline
    A & 25  &  5 & isotropic          & \nodata  & \nodata & $-3.5$ & 0.1 & 1 \\
    B & 38  &  5 & isotropic          & \nodata  & \nodata & $-3.5$ & 0.1 & 1 \\
    C & 50  &  5 & isotropic          & \nodata  & \nodata & $-3.5$ & 0.1 & 1 \\
    D & 100 & 20 & isotropic          & \nodata  & \nodata & $-3.5$ & 0.1 & 1 \\
    E & 50  &  5 & sunward            & 225 & \nodata & $-3.5$ & 0.1 & 1 \\
    F & 50  &  5 & sunward            & 180 & \nodata & $-3.5$ & 0.1 & 1 \\
    G & 50  &  5 & sunward            &  90 & \nodata & $-3.5$ & 0.1 & 1 \\
    H & 60  &  6 & $\cos^2(\theta/2)$ & \nodata  & \nodata & $-3.5$ & 0.1 & 3 \\
    I & 60  &  6 & $\cos^2(\theta/2)$ & \nodata  & \nodata & $-3.6$ & 0.1 & 3 \\
    J & 60  &  6 & $\cos^2(\theta/2)$ & \nodata  & \nodata & $-3.7$ & 0.1 & 3 \\
    K & 60  &  6 & $\cos^2(\theta/2)$ & \nodata  & \nodata & $-3.6$ & 0.1 & 1 \\
    L & 60  &  6 & $\cos^2(\theta/2)$ & \nodata  & \nodata & $-3.6$ & 0.1 & 10 \\
    M & 60  &  6 & $\cos^2(\theta/2)$ & \nodata  & \nodata & $-3.6$ & 0.5 & 3 \\
    N & 60  &  6 & isotropic          & \nodata  & \nodata & $-3.6$ & 0.5 & 3 \\
    \enddata
    \tablecomments{Parameter descriptions: ($s_1$) mean ejection speed for $a=1$~\micron{}; ($\sigma$) Gaussian width of speed scale factor; ($\hat{v}$) ejection direction, coordinates are J2000 RA and Dec., and $\theta$ in the outburst models is measured from RA, Dec. = $(140\degr, 30\degr)$; ($w$) ejection cone opening angle; ($k$) grain differential size distribution power-law slope ($dn/da\propto{}a^k$) at the nucleus surface; ($a_{\mathrm{min}}$, $a_{\mathrm{max}}$) minimum and maximum grain radii.}
  \end{deluxetable*}

  Ejection anisotropies are required to reproduce the comet's lightcurve and morphology, and the models with production towards (RA, Dec) = ($210$\degr, $-10$\degr) are in best agreement with both the comet's lightcurve and morphology.  This ejection source is illustrated with Models (A--D) in Figs.~\ref{fig:quiescent-lightcurve} and \ref{fig:ztf}.  We consider model (A) to be the best overall.  Model (B) is the same as model (A), but with $k$ decreased from --2 to --6 to show the insensitivity to this parameter.  Model (C) is the best lightcurve model for each of $a_\mathrm{max}=1$~cm and $s_1=50$~\mps.  The addition of the larger dust particles produces narrower and longer tails.  Model (D) is separately the best lightcurve model for $N=-3.5$ and $s_1=100$~\mps.  The higher expansion speed produces a wider pre-opposition tail and a more symmetric post-opposition morphology.  Compared to the ($210$\degr, $-10$\degr) source direction, sunward ejection models (E--G) have poorer agreement with the images (compare the pre-opposition tail orientation, and post-opposition bias towards position angles +90\degr{} from the Sun direction).  Finally, the isotropic ejection model is much too broad, even for the low ejection speed $s_1=38$~\mps.

  No model precisely matched the image data, but model (A) was the most successful.  Some improvements may be possible, e.g., by adding some larger particles or decreasing $w$ to thin out the tail, or by moving the dust ejection cone away from ($210$\degr, $-10$\degr) for a better post-opposition brightness.  Given the large number of free parameters, and the likelihood of partial degeneracies between some of them, we take model (A) as final, but assume that a fully optimized model will be found with the values $s_1$ = 38 to 50~\mps, $w$ = 45 to 90\degr, $k$ = $-3$ to $-1$, $N$ = $-3.2$ to $-3.4$, and $a_\mathrm{max}$ = 1 to 10~mm.

  The dust mass-loss rate at perihelion, $Q_P$, is $2.7\pm0.6$~\kgps, calculated from the mean and standard deviation of the mass-loss rates after varying parameters $s_1$, $w$, $N$, and $k$ in combination using their likely ranges given above.  For $a_\mathrm{max}=1$~cm, $Q_P=10\pm3$~\kgps.

  \subsection{Outburst}\label{sec:outburst}

  \subsubsection{Lightcurve}\label{sec:outburst-lightcurve}

\begin{figure*}
    \centering
    \includegraphics[width=\textwidth]{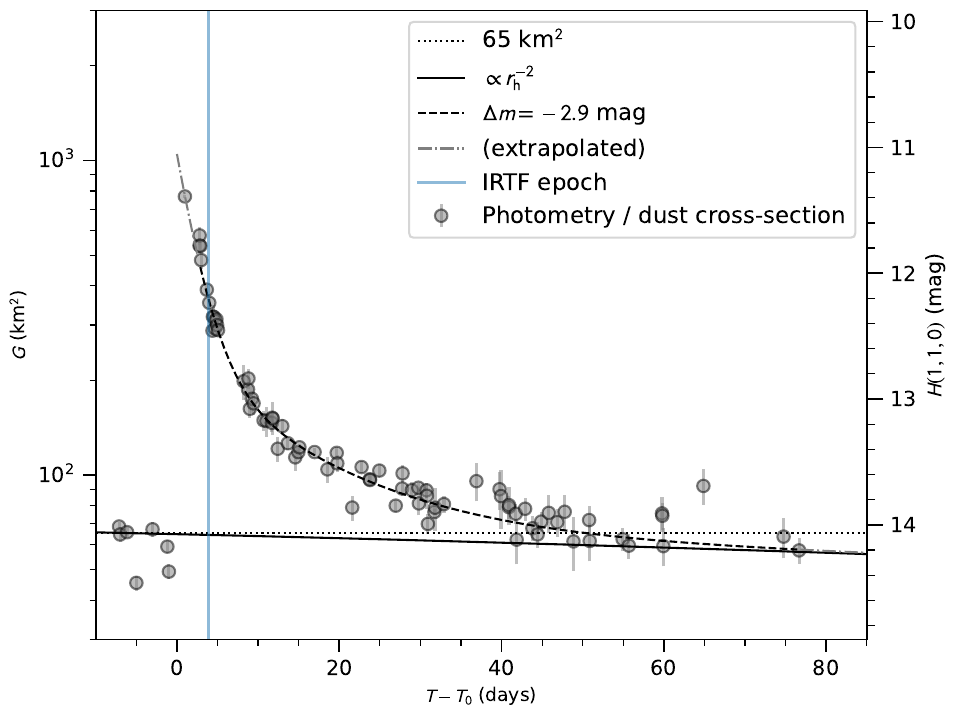}
    \caption{Variation of dust geometric cross section, $G$, and absolute magnitude, $H$, with time within a 10,000-km radius aperture during the December 2021 outburst.  The dotted line corresponds to a constant dust cross section of 65 km$^2$. Our adopted baseline coma brightness is shown as a solid black line, and the best-fit two-exponential outburst model is presented as a dashed line.  The portions of the exponential that have been extrapolated beyond the fitted data are shown as dashed-dotted lines.  A vertical line marks the observation time of the IRTF SpeX spectrum.}
    \label{fig:outburst}
  \end{figure*}

  With an approximate model for the quiescent coma, we now turn to the lightcurve and dynamics of the December outburst.  During an outburst the assumption of a constant outflow is violated, therefore the benefits of using the \afr{} quantity are reduced.  Instead, we convert observed spectral flux density into effective geometric cross section, $G$ (units of area),
  \begin{equation}
    G = \frac{\pi \Delta^2}{A_p \Phi(\phi)}\frac{F_\nu}{S_\nu(\rh)},
  \end{equation}
  where $A_p$ is geometric albedo (dimensionless), and $\Phi(\phi)$ is the (Schleicher-Marcus) phase function (dimensionless) evaluated at phase angle $\phi$.  For $A_p$, we assume 0.040 at 0.55~\micron{} and extrapolate to the $g$-, $r$-, and $o$-bands (0.036, 0.043, and 0.046, respectively), using our measured coma color.  The results are listed in Table~\ref{tab:obs}.  The cross-sections are shown in Fig.~\ref{fig:outburst}.  Cross-section may be converted to absolute magnitude, $H$, i.e., the apparent brightness of the dust as seen by a Sun-based observer at a distance of 1~au:
  \begin{equation}
    H = m - 5 \log_{10}(\rh \Delta) = -2.5 \log_{10}\left( \frac{A_p G }{\pi (1\mathrm{~au})^2} \right) + m_\sun
  \end{equation}

  The total dust cross-section falls with time to below the pre-outburst level of 65~km$^2$.  In accordance with our best quiescent model, we scale the pre-outburst baseline with $\rh^{-2}$ to define a baseline trend that is in agreement with the last ZTF and TRAPPIST data at +76 days.  The trend introduces a 0.14~mag decrease in brightness over the 80-day post-outburst period.

  We initially fit the outburst cross-sectional areas with a single exponential, $G e^{-t/\tau}$, but with poor results ($G=749\pm9$~km, $\tau=4.8\pm0.1$~days, reduced $\chi^2=20$).  The sum of two exponential functions provided improved fits (reduced $\chi^2=2.1$), with a total outburst strength of $\Delta m=-2.9$~mag, and is shown in Figs.~\ref{fig:lightcurve} and \ref{fig:outburst}.  The peak cross-sections and decay timescales are $G_1=822\pm16$~km$^2$, $\tau_1=2.5\pm0.1$~days, and $G_2=165\pm12$~km$^2$, $\tau_2=14.8\pm0.8$~days for the two components.  The large reduced $\chi^2$ value is at least partially a result of unaccounted systematic uncertainties in the photometry.  In addition, these fits assumed our baseline trend accurately represents the comet's quiescent activity, but this is not necessarily the case.  Moreover, the uncertainty on the timing of the outburst has a substantial impact on the cross sections and timescales ($\pm$40\%).  Instead, the total cross section should be measured directly from the first post-outburst ATLAS image.  We take our two-component fit as a notional description of the lightcurve shape, but do not interpret the two components literally.

  \subsubsection{Morphology}\label{sec:outburst-morphology}
  We returned to our dynamical model to reproduce the morphology of the LDT data at 5~days after the outburst (Fig.~\ref{fig:morph}d).  Outburst model parameter sets discussed in the text are listed in Table~\ref{tab:model}, but many more simulations were made to explore the parameter space.  All models assume an impulsive event (i.e., delta function in time) at our nominal outburst date.

\begin{figure*}
    \centering
    \includegraphics[width=\textwidth]{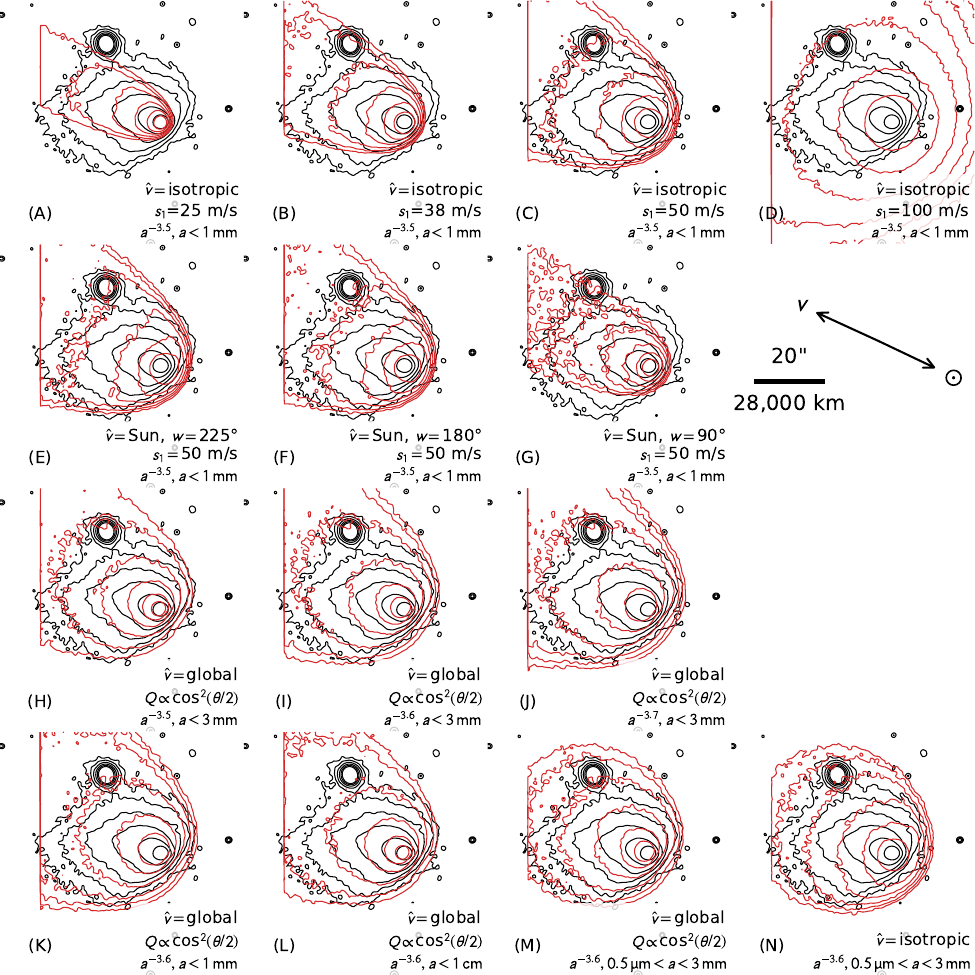}
    \caption{Morphology of the LDT image taken 2018 December 16 and example outburst models.  The data and models are shown as black and red isophotal contours, respectively, spaced at factors of two intervals from $2^{-3}$ to $2^{-8}$.  Panel labels correspond to the outburst models in Table~\ref{tab:model}.  All models use $s_1=60$~\mps, 0.1~\micron{}$<a<$1~mm, and $dn/da\propto a^{-3.5}$, unless noted.  The images were smoothed with a $\sigma=0\farcs5$ Gaussian kernel.  Image scale and orientation are given in the upper right.  Each row signifies the variation of a different parameter: (A)--(D) varies ejection speed, (E)--(G) varies the opening angle of a sunward cone, (H)--(J) varies the power-law slope on the grain size distribution, and (K)--(N) varies grain size range.  Model (M) better represents the compactness of the ejecta and the flattened sunward distribution (see Section~\ref{sec:outburst-morphology} for a discussion), although additional asymmetries and perhaps the addition of the quiescent activity may be necessary to account for the elongation of the isophotes towards PA$\sim90$\degr.}
    \label{fig:outburst-morph}
  \end{figure*}

  The ejecta has an edge near 6000--7000~km in the sunward direction.  Initial tests covering a range of ejection speeds ($s_1=25-150$~\mps) indicated that lower values, $\lesssim$50~\mps, were necessary to make the sunward extent of the dust in agreement with the data (see models A--D in Fig.~\ref{fig:outburst-morph}).

  In Fig.~\ref{fig:outburst-morph}, we show $s_1$=50~\mps{} models with wide ejecta cones directed toward the Sun using $w$=225, 180, and 90\degr{}, labeled as models (E), (F), and (G), respectively.  The models produced parabolic edges in the sunward direction and photometric contours flared in the directions perpendicular to the sun vector.   In contrast, the image has a flattened shape in the sunward direction and contours elongated away from the Sun.  A broad sunward ejecta cone does not reproduce the ejecta distribution.

  Isotropic model runs were again split into a series of active areas evenly distributed in RA and Dec, this time 50 sources in total ($\approx27\degr$ spacing).  Some sources on the illuminated hemisphere with broad opening angles appeared to have better agreement with the data.  No model reproduced the tail-ward curvature.  In the sunward direction, all models presented segments of the paraboloid shape of the isotropic model, i.e., none approximated the flattened shape.

\begin{figure*}
    \includegraphics[width=\textwidth]{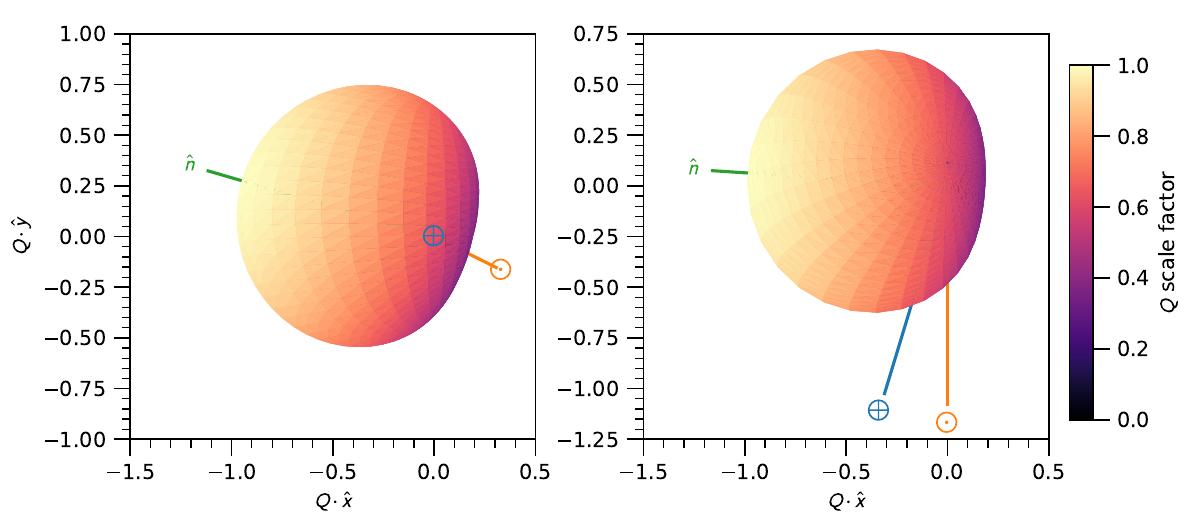}
    \caption{Dust production scaling function ($Q$, dimensionless) that produces an ejecta morphology in best agreement with our first LDT image.  The production rate follows $\cos^2(\theta/2)$, where $\theta$ is measured from RA, Dec.\ = (140\degr, 30\degr), shown here as the vector $\hat{n}$.  The comet-Sun ($\sun$) and comet-observer ($\oplus$) vectors are also shown.  (Left) The view from the Earth ($\hat{y}$ is equatorial north, $\hat{x}$ is west).  (Right) Top-down view of the Sun-comet plane ($\hat{y}$ is the Sun-comet unit vector, $\hat{x}$ is perpendicular to the Sun direction in the comet's orbital plane).}
    \label{fig:outburst-asymm}
  \end{figure*}

  Based on the initial models (A--G), we re-considered the assumption that the ejecta parameters are uniform within the ejecta cone, and assumed anisotropic dust production, $Q$, as a function of angle, $\theta$, measured with respect to a nominal vector.  We experimented with $Q\propto \cos(\theta/2)$, $\cos^2(\theta/2)$, and $\cos^4(\theta/2)$.  These functions eject dust in all directions (global production) but are peaked in the direction of a nominal vector.  We will discuss the physics behind this scenario in Section~\ref{sec:discussion}.  An initial exploration of nominal vectors indicated that vectors near the image plane were better at reproducing the anisotropy.  Because the comet-observer and comet-Sun vectors only differ by 18\degr, we limited our final search to vectors perpendicular to the comet-Sun direction, i.e., near the solar terminator for a spherical nucleus.  Production weighted by $\cos^2(\theta/2)$ measured from the vector pointing towards RA, Dec.\ = (140\degr, 30\degr) better represented the LDT image under consideration.  This production function is visualized in Fig.~\ref{fig:outburst-asymm}.  Small, $\sim10$\degr{} rotations of the vector in the plane perpendicular to the Sun also rotate the sunward asymmetry (not shown), and we consider this the approximate limit of our characterization of the outburst direction.

  Models (H--J) in Fig.~\ref{fig:outburst-morph} use the $\cos^2(\theta/2)$ distribution to show the effects of varying the particle size distribution power-law slopes.  Three models are presented for particles up to 3~mm in radius with slopes $-3.5$, $-3.6$, and $-3.7$, models (H), (I), and (J), respectively.  Increasing the slope index from $-3.7$ to $-3.5$ concentrates more scattered light near the nucleus.

  Next, we further varied the grain size distribution to show the effects when the large particle radius limit is varied.  Models (K), (I), and (L) correspond to $a_\mathrm{max}$=1~mm, 3~mm, and 1~cm, respectively.  Including particles up to 1~cm in radius adds more flux to the central isophot and produces better morphologies.  However, varying the grain size distribution has substantial consequences on the lightcurve that will be described later.

  Finally, we examined the effect of varying the minimum grain size from our nominal 0.1~\micron{} to 2~\micron.  Model (M) of Fig.~\ref{fig:outburst-morph} shows the $a_\mathrm{min}=0.5$~\micron{} model.  The compact shape and flattened sunward distribution of the LDT image is better accounted for by this model, but the agreement with inner contours is somewhat degraded (compare with models (H) and (L)).  Additional ejection asymmetries are likely necessary to fully reproduce the data, and possibly considerations for the tail from the comet's quiescent behavior.  Regardless, further information on the model parameters will be gained from comparisons to the lightcurve in the next section.

  We ran additional simulations for the second LDT image, taken $32\pm1$~days after the outburst.  They predict a ring or crescent morphology near the nucleus, the diameter of which depends on the ejection speed and size of the largest particles.  For the $s_1=60$~\mps{} models with $a<3$~mm, the crescent is 2\farcs4 in diameter.  We find no evidence for such a ring in the data.  However, its presence would be affected by the 2\farcs1 seeing, and our lightcurve analysis suggests the ejecta is only 30\% of the flux in a 6\arcsec{} radius aperture.  There are no obvious signatures of any ejecta in our LDT image.  Without a way to identify the ejecta morphology, we avoid any further model refinement based on this image.

  \subsubsection{Final models}\label{sec:outburst-models}

  With ejection speed and direction estimated, we simulated the outburst lightcurve.  Ten low-resolution ($10^6$ particle) simulations were generated from $T_0+0$ to $+80$~days.  Simulated photometry was derived from a 10,000~km radius aperture and calibrated to the lightcurve at $T_0+5$~days.  A selection of model lightcurves, presented as geometric cross sectional area, is shown in Fig.~\ref{fig:outburst}.  In contrast with the morphology, the lightcurve is more forgiving.  Similar quality fits can be obtained with other ejection speeds (e.g., $s_1=50$~\mps{} versus 60~\mps), grain size distributions, or even with an isotropic coma (e.g., model (N)).

  \begin{figure*}
    \includegraphics[width=\textwidth]{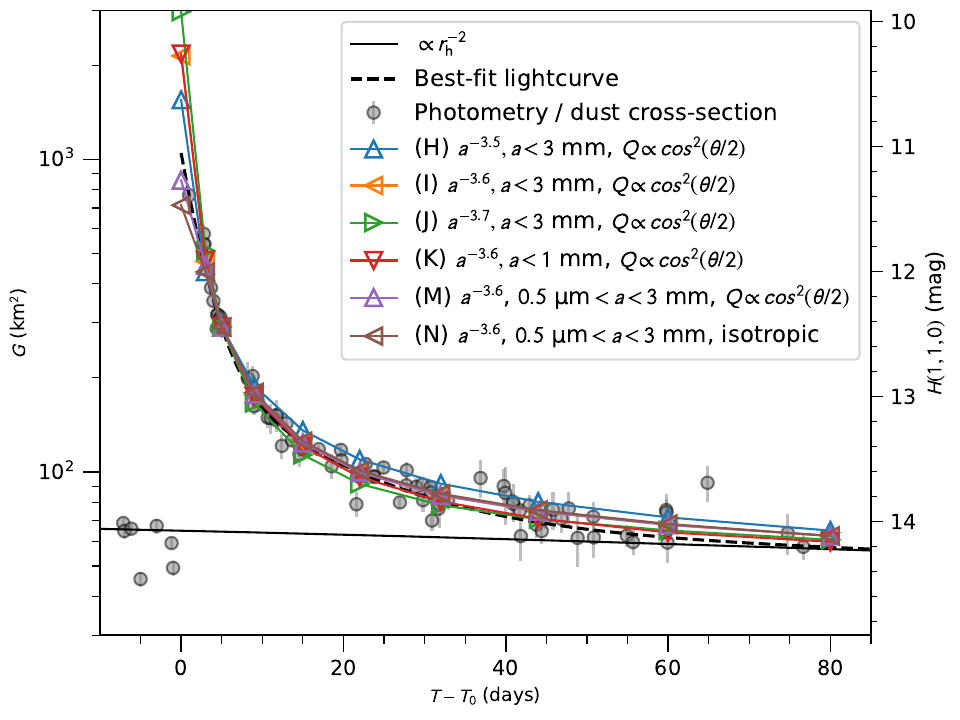}
    \caption{Same as Fig.~\ref{fig:outburst}, but including model outburst lightcurves.  Triangles represent a subset of dynamical models tested (Table~\ref{tab:model}).  Our adopted baseline coma brightness is shown as a solid black line, and the best-fit two-exponential outburst model is presented as a dashed line.  Model (M) best represents all aspects of the data, but model (N) is still a fair representation of the lightcurve.}
    \label{fig:outburst-models}
  \end{figure*}

  The consequences of varying the grain size distribution are illustrated in Fig.~\ref{fig:outburst-models}.  Initially, the smallest dust grains dominate the scattered light within the photometric aperture (10,000-km radius).  These are also the fastest moving grains, based on our adopted size-speed relationship ($s\propto a^{-1/2}$).  As a result, the strength of the outburst peak and the initial slope are sensitive to the grain size distribution.  The smallest grains leave the aperture first, and the slower expansion of the larger grains produces the long tail in the lightcurve's quasi-exponential decay.

  Models (H), (I), and (J) are based on differential grain size distributions proportional to $a^{-3.5}$, $a^{-3.6}$, and $a^{-3.7}$.  Model (J), $a^{-3.7}$, with the greatest proportion of small grains has the strongest peak brightness and most rapidly falling lightcurve.  Of the three models (H), $a^{-3.5}$, with the weaker peak brightness more closely follows the post-outburst trend.  But this improved agreement comes with a cost at later post-outburst times.  The greater proportion of large particles in model (H) slows the long-term fading which over-predicts the lightcurve starting about 8~days post-outburst.  Despite the strong initial fading, model (I), $a^{-3.6}$, is in good agreement with the lightcurve at $>3$~days post-outburst.

  In Section \ref{sec:outburst-morphology}, a small particle cutoff improved the agreement with the outburst morphology.  In Fig.~\ref{fig:outburst-models}, we show model (M) which is the same as model (I), $a^{-3.6}$, but with the minimum grain size increased from 0.1 to 0.5~\micron.  This change reduced the peak brightness and brought the model lightcurve into good agreement with the ATLAS photometry from the discovery image ($T-T_0=0.98$~days).  The remainder of the model lightcurve at $T_0>5$~days was essentially unchanged.  Model (M) best represents the total lightcurve.

  The largest particles dominate the brightness at later times.  They move more slowly and can linger in the photometric aperture for months.  The presence of particles larger than 3~mm cause the late lightcurve ($>T_0+40$~days) to be too bright.  However, this interpretation relies on our adopted quiescent brightness model, which follows \rh$^{-2}$.  Larger particles can be accommodated with a stronger heliocentric distance slope.  Without an independent measurement of the activity for this portion of the orbit (which may be subtly affected by coma asymmetries), we conclude the large particle cutoff cannot be confidently assessed from the lightcurve alone.

  Finally, an estimate of the ejecta dust mass may be made with the total brightness of the comet in the ATLAS discovery image: $o=15.31\pm0.02$~mag corresponds to $G=819\pm15$~km$^2$ and $M=(3\pm1)\times10^7$~kg.  The 30\% relative uncertainty in our outburst mass retrieval is dominated by the abundance of large particles.  Whether we vary the size distribution slope by $\pm$0.1 ($-3.5$ to $-3.7$), or the large grain limit by a factor of 10 (1~mm to 1~cm), the mass uncertainty range is the same.  Assuming the same grain size distribution, the small outbursts ejected $\sim10^5$~kg of dust.

  \section{Discussion} \label{sec:discussion}

  Water ice has been directly observed in near-infrared spectra following outbursts of comets 17P/Holmes, P/2010 H2 (Vales), and 29P/Schwassmann-Wachmann 1 and in the material excavated by the Deep Impact experiment at comet 9P/Tempel~1 \citep{sunshine07,yang09-holmes,yang10-iauc,protopapa21-atel14961}.  No near-infrared water-ice spectral features, including the strong 3-\micron{} band, were seen in one (natural) outburst of comet 9P/Tempel~1 and two outbursts of comet 67P/Churyumov-Gerasimenko \citep{moretto17-tempel1,bockelee-morvan17-outbursts}.  For comet 243P, we found no evidence for water-ice absorption features in our near-IR spectrum of the outburst ejecta, with an upper-limit of 3\% on the 2-\micron{} absorption band depth.  Either ice was never ejected in these latter events, or water-ice grains were ejected but undetectable given the circumstances of the observation.  The ice may be made hidden by low spectroscopic signal-to-noise ratios, a paucity of water ice in the ejecta (i.e., strongly dust-dominated), dynamical escape of icy grains from the spectroscopic aperture, or water ice sublimation over the time since the outburst occurred.  For 243P's December outburst, we found that the signal-to-noise ratio was sufficient to detect ice at levels previously seen in cometary spectra.  We consider the remaining possibilities in turn, then discuss the mass, dynamics, and specific energy of the outburst.

  \subsection{Were any water-ice grains ejected by the outburst?}
  We first consider the ice content of the ejecta.  Cometary surfaces are devolatilized by insolation, but the devolatilization depth is likely shallow compared to the depth of excavation by 243P's large outburst.  Assuming a nuclear bulk density of 500~kg~\inv[3]{m}, $3\times10^7$~kg of material corresponds to a hemispherical crater with a depth of 30~m (median depth by mass is 10~m).  The mass and excavation depth are similar to those of the \di{} experiment, which liberated of order $10^7$~kg of material from the nucleus of comet 9P/Tempel~1 \citep{ahearn05, ahearn08}, excavating a transient crater 10--50~m deep \citep{schultz13, richardson13}.  The \di{} spacecraft observed water ice in the ejecta \citep[2.0, and 3.1~\micron{} features;][]{sunshine07}, and telescopic observations of the event showed other evidence of water ice \citep{kueppers05,schulz06-di,harrington07-di,fernandez07,gicquel12}.  If the sub-surface composition of 243P is the same as that of 9P/Tempel~1, then we should expect icy grains to have been ejected.

  A comparison to the nucleus of comet 67P leads us to the same conclusion.  At comet 67P/Churyumov-Gerasimenko, water-ice sublimation occurs within $\sim1$~cm of the surface \citep{schloerb15}.  If the sub-surface of 243P is similar, then it is unlikely that excavation depths $\sim$10~m would be devoid of ice.

  Despite the near-surface availability of water ice, near-infrared features attributable to water ice grains were notably absent from three well-studied outbursts of comets 9P and 67P \citep{moretto17-tempel1, bockelee-morvan17-outbursts}.  In particular, the strong 3-\micron{} water ice absorption feature was covered by the data sets considered by both investigations, yet the spectra only yielded upper limits on the band depth: 18\% for the outburst of 9P (M.~Moretto, private communication) and 10\% for the two outbursts of 67P \citep{bockelee-morvan17-outbursts}.  Compare these results to those of comet 103P presented in Section~\ref{sec:results:spectroscopy}.  The icy Box B spectrum has a 3-\micron{} band depth of 20\% (Fig.~\ref{fig:di-spectra}), corresponding to a 5\% abundance by area based on a model of 0.5~\micron{} radius water ice grains.  Thus, the outbursts of 9P and 67P do not necessarily lack water ice altogether, but may just have a lower abundance than seen in other events or quiescent comae.  In fact, evidence for water ice was seen in some of comet 67P's outbursts, but at near-UV wavelengths by the Rosetta, which may indicate a population of ice grains smaller than 0.1~\micron{} \citep[][and references therein]{noonan21-outbursts}.

  Altogether, the evidence is in favor of the excavation of water ice by the large outburst of comet 243P.  What remains for us to understand is what are the circumstances that would prevent us from detecting it in our data, other than low abundance.

  \subsection{What are the dynamics of small icy grains?}
  Water-ice grains observed in cometary comae are on the order of $\sim$0.1--1~\micron{} \citep[e.g.,][]{kawakita04-ice,yang09-holmes,protopapa18-catalina}.  The dynamics of small icy grains could have moved them outside of our spectroscopic slit by the time of our observations.  In Section~\ref{sec:outburst}, we presented a dynamical model of the outburst ejecta based on the optical properties of low-albedo dust (amorphous carbon).  The radiation pressure efficiencies calculated by Mie theory for water ice are about a factor of 3 smaller than those of amorphous carbon, for $a\geq0.5$~\micron.  Thus, water-ice grains feel less acceleration from sunlight than the amorphous carbon grains in our model.  However, it is primarily the ejection speed that dominates their initial dispersion on small angular scales.  We used our dynamical model to estimate if small water-ice grains could still be present in the SpeX slit 4~days after the outburst.

  We calculated the areal fraction of the ejecta remaining in a 2\farcs1 radius (3000~km) aperture as a function of size, based on our best dynamical model (M).  Fine dust dominates at early epochs.  Particles smaller than 3~\micron{} in radius account for 67\% of the total geometric cross section.  By the time of the SpeX spectrum, this fraction has dropped to 1--2\%.  Thus, the small end of the grain size distribution has been depleted by a factor of $\sim$50, relative to the remaining dust.  We can speculate that if the original composition of the ejecta contained water ice with properties similar to other comets, then the initial 2-\micron{} band-depth of $\sim$10\% would likely be diminished well below our observed 2-\micron{} band-depth upper-limit of 3\%.  The dynamical depletion of small grains within the SpeX slit width (0\farcs8) suggests they may not be numerous enough to be detected spectroscopically.

  We repeated this exercise for the LDT color maps (Fig.~\ref{fig:color-map}) by considering the population of grains located 30,000 km down the ejecta plume.  Grains between 1 and 3~\micron{} in size account for 65\% of the cross-sectional area at the time of the LDT observation.  This dynamical analysis suggests that micrometer sized water ice grains should still be present in the LDT color map, 5 days after the outburst.

  \subsection{What are the sublimation lifetimes of small icy grains?}
  Finally, we consider the lifetime of water ice at 2.55~au with an ice sublimation model.  We use the grain sublimation code of \citet{kelley23-grains2-v0.5.0} previously used by \citet{protopapa18-catalina, protopapa21-wirtanen} to estimate the dust fraction of water ice bearing grains in the coma of comets C/2013 US$_{10}$ (Catalina) and 46P/Wirtanen.  The model balances insolation with energy lost from sublimation and thermal radiation, and considers additional losses due to solar wind sputtering \citep{mukai81-sputtering}.  Water ice is described with the optical constants of \citet{warren08}, the latent heat of sublimation from \citet{delsemme71}, and the vapor pressure equation of \citet{lichtenegger91}.  Dust-ice aggregates are created by mixing optical constants with the effective medium approximation of \citet{bruggeman1935}, and we assume the grain dust-to-ice ratio is constant with time, even as the grain size decreases due to sublimation (i.e., a homogeneous dust and ice mixture).  We define the grain lifetime as the time for a grain to sublimate down to a 10~nm radius.

  \citet{protopapa18-catalina} detected 1.5- and 2.0-\micron{} water ice features in the coma of comet C/2013 US$_{10}$ (Catalina).  Ice was seen when the comet was at 5.8, 5.0, and 3.9~au from the Sun, but not at 2.3~au.  They argued that the lack of ice features at 2.3~au was an indication that the coma ice could not be pure, and was instead in contact with some low-albedo material.  A model with an icy grain radius of 0.6~\micron{} and a dust fraction of $\sim$0.5\% by volume could explain the water-ice absorption bands at $\geq$3.9~au and the lack of the bands at $\leq$2.3~au.  We will use the ice properties of the coma of comet Catalina as a template for water ice around comet 243P.

\begin{figure*}
    \centering
    \includegraphics[width=\textwidth]{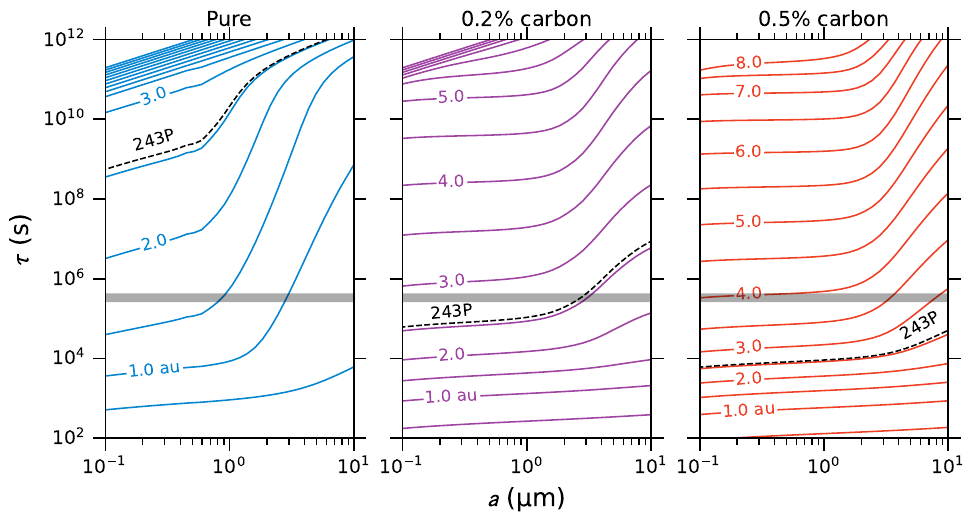}
    \caption{Water-ice grain lifetimes, i.e., time to complete sublimation, versus grain size for pure and dirty grains (0.0, 0.2, and 0.5\% amorphous carbon by volume). Lines of constant heliocentric distance are shown as solid lines from 0.5 to 8.0~au in 0.5~au steps.   A dashed line indicates grain lifetimes specific to the circumstances of the 243P/NEAT outburst of December 2021 at 2.55~au from the Sun.  A horizontal bar indicates the post-outburst time for our near-infrared spectrum of 243P.  Grains on the 243P line with lifetimes below the shaded region would have sublimated away before being observed.}
    \label{fig:sublimation}
  \end{figure*}

  In Fig.~\ref{fig:sublimation}, we present sublimation lifetime calculations for heliocentric distances from 0.5 to 8~au and grain radii between 0.1 and 10~\micron.  As found by other investigators \citep[e.g.,][]{hanner81-albedo}, pure water-ice grains have extremely long lifetimes, $10^{10}$~s at 2.55~au for micrometer-sized grains.  Such grains would have been present in the ejecta at the time of our IRTF spectrum and LDT color map ((3--4)$\times10^5$~s post-outburst).  Included in Fig.~\ref{fig:sublimation} are calculations for icy grains containing a small amount of dust, 0.2\% and 0.5\% amorphous carbon by volume.  The time to total grain sublimation is significantly reduced to $10^5$ and $10^4$~s, respectively, for ice at 2.55~au.  Icy grains with just a small amount of low albedo dust are likely to have sublimated before the time of our IRTF spectrum and LDT color map.

  \subsection{Outburst mass, dynamics, and specific energy}
  The large outburst of 243P produced a similar amount of dust as the quiescent activity observed during our 251-day observational data set.  The quiescent production rate for our best-fit model yields a total mass loss of $5.5\times10^7$~kg over the observed 251~day period.  Compare this value to the total mass of the large outburst, $(3\pm1)\times10^7$.  Given that our observations span a portion of the comet's 7.5~year orbital period, the outburst accounts for $\lesssim55$\% of the orbital mass loss around the 2018 perihelion.  In absolute mass terms, \citet{ishiguro16-outbursts} found that the Jupiter-family comet population produces at least one similarly sized event or larger every year.

  We found that the outburst morphology could be approximated with an initial outflow following a $\cos^2(\theta/2)$ distribution.  Although an accurate representation of all aspects of the data was not achieved, the adopted best model, (M), did yield a good approximation to the ejecta in the projected sunward direction.  Assuming that the outburst ejecta freely expands from a point on a spherical nucleus, a $\cos^2$ distribution, which ejects dust in all directions, would not be possible as material can only move away from the nucleus and not through it.  Perhaps a small amount of back-flow could be produced if the site of the outburst were at the tip of a high ridge or pinnacle, but likely not enough to account for the $\cos^2$ distribution.

  If the outburst event were to occur in the coma, such as by a disintegrating large chunk of nucleus, then the appearance of widespread or global production could be possible.  The 243P outburst mass corresponds to a 24-m radius sphere for a density of 500~kg~m$^{-3}$.  The spontaneous disintegration of such an object could occur due to rotationally induced fragmentation.  Based on the work of \citet{jewitt20-borisov}, the spin-up timescales for an icy nucleus of this size at 2.5~au range from $\sim$3~days to $\sim$8~months, depending on the cohesive strength of the nucleus and the driving volatile, here, CO versus \water, respectively.  In this scenario the CO-driven case is less likely, as no fragment producing event was seen in the weeks before the large outburst (Fig.~\ref{fig:lightcurve}), and CO tends to be depleted in the Jupiter-family comet population \citep{harrington-pinto22-coco2}.  Instead, water-driven torques may be plausible, if the fragment were produced during an unobserved event (e.g., before June 2018), or during the September 2018 mini-outburst.  This scenario additionally requires that the fragment be kept in proximity to the nucleus over long timescales (i.e., in orbit), or else the outburst would have been displaced from the comet's ephemeris position.  However, this scenario does not naturally explain why the ejecta moves in a preferred direction ($\sim$90\degr{} from the comet-Sun direction toward RA, Dec. = 140\degr, +30\degr) after disintegration.

  \citet{ishiguro16-outbursts} compared three cometary outbursts: $\sim10^8$~kg outbursts of comet 15P/Finlay and 332P/Ikeya-Murakami, and the $\sim10^{11}$~kg mega-outburst of comet 17P/Holmes in 2007.  The specific energy (kinetic energy per unit mass) estimates of the events were similar, $10^4$~J~kg$^{-1}$ for 15P and 332P, and $10^4-10^5$~J~kg$^{-1}$ for 17P.  Similarly, \citet{ye15-15p,ye24-15p-erratum} estimated a specific energy of 0.3--$2\times10^5$~J~kg$^{-1}$ and a mass loss of $(2-5)\times10^8$~kg for two outbursts of 15P.  This led \citet{ishiguro16-outbursts} to suggest that the outburst mechanisms of these three comets were similar, despite the four orders of magnitude range in ejected mass.  \citet{li11-holmes} found that the specific energy produced by the exothermic phase change of amorphous water ice to crystalline water ice would be of order $10^5$~J~kg$^{-1}$, and concluded it was a plausible outburst mechanism for 17P/Holmes.  \citet{ishiguro14-332p} and \citet{ishiguro16-holmes} concluded the same for the 332P and 15P outbursts.

  In contrast, \citet{jewitt20-p2010h2} estimated 220~J~kg$^{-1}$ for a major outburst, $1\times10^9$~kg, of comet P/2010 H2 (Vales).  This is three orders of magnitude smaller than the specific energy expected for the amorphous to crystalline water ice phase change.  However, they concluded that the exothermic phase change could have driven the outburst, provided there was an inefficient conversion from crystallization energy to ejecta kinetic energy.  No reasoning or model for this inefficient conversion has been proposed.  In a near-infrared spectrum of this comet's outburst ejecta, \citet{yang10-iauc} identified a 1.65~\micron{} absorption feature as a signature of crystalline water ice.  Yet, the crystallization timescale at 3.1~au is 55~s \citep{prialnik24-comets3}, so one cannot conclude whether or not the crystallization occurred in the nucleus or coma.

  \citet{jewitt20-p2010h2} considered driving mechanisms other than amorphous water ice for the outburst of comet P/2010 H2.  They found that expelled debris from a cliff collapse could easily provide the surface area needed for CO ice sublimation or amorphous water ice conversion to drive the outburst based on the measured outburst mass loss rate.  They ruled out an impact scenario due to low collision probabilities, and they ruled out several low energy processes (rotational disruption, thermal fracture, electrostatic forces) as they alone cannot achieve the high expansion speeds of the dust.

  The large outburst of 243P produced $\sim10^7$~kg of material with a kinetic energy per mass of $76$~J~kg$^{-1}$, based on outburst model (M).  Varying $s_1$ from 50 to 70~\mps, $a_\mathrm{max}$ from 1 to 10~mm, and the differential grain size distribution power-law slope from $-3.5$ to $-3.7$ yields values from 18 to 240~J~kg$^{-1}$.

  A comparison between the outbursts of the above five comets suggests a grouping in their energies and the physical processes at work: a low energy process produced the outbursts of comets P/2010 H2 and 243P with specific energies $\sim10^1-10^2$~J~\inv{kg}, versus the high energy process that produced the outbursts of comets 15P, 17P, and 332P with specific energies $\sim10^4-10^5$~J~\inv{kg}.  Given that amorphous water ice crystallization has been proposed to explain these events, we consider what difference may lead to the lower specific energies for comets P/2010 H2 and 243P.

  Gases more volatile than water appear to be necessary for outbursts driven by amorphous water ice crystallization, as this exothermic reaction may not produce enough water gas to eject large amounts of dust \citep[][and references therein]{prialnik24-comets3}.  Instead, gases, such as \coo{} or CO, trapped within the water ice would be liberated by the phase transition.  These super-volatile gases would be able to eject dust from the nucleus.  We propose that, if the crystallization of water ice was responsible for the aforementioned outbursts then the cause for the range in specific energies may be due to a difference in the relative abundance of trapped volatile gases.  Specifically, the lower specific energies of 243P and P/2010~H2 would indicate a lower abundance of \coo{} and/or CO relative to the converted water ice, as compared to comets 15P, 17P, and 332P.

  The kinetic energy per unit mass of the outburst ejecta is similar to that of the quiescent activity, which is 13~J~kg$^{-1}$ for quiescent model (A).  Varying $s_1$ from 38 to 50~\mps, $a_\mathrm{max}$ from 1 to 10~mm, and the differential grain size distribution power-law slope from $-3.2$ to $-3.4$ yields values from 2 to 38~J~kg$^{-1}$.  Rather than requiring an efficient conversion of energy from amorphous water ice crystallization, the outburst could instead be viewed as a temporarily high mass-loss rate, $>180$~\kgps, driven by typical cometary sublimation.  The lower limit on the mass-loss rate is more typical of bright active comets near 1~au (a water production rate of $1\times10^{28}$ molecules~s$^{-1}$ is 300~\kgps).  Enhanced activity could be initiated by cliff collapse or a landslide \citep[e.g.,][]{agarwal17-outburst,steckloff18}.  However, the activity must be quenched on rapid timescales, by volatile depletion or insulation from the Sun.  This is unlike the scenario observed at, e.g., comet 240P/NEAT, where a rapid onset of new activity appeared to be sustained for several months \citep{kelley19-240p}.  Measurements of the coma gases, especially the primary species in comets (\water, \coo, and CO), before, during, and after the outburst would have benefited this study, allowing us to determine if the gas coma was similarly enhanced by the outburst, or if particular gas species, e.g., \coo, may have driven the event \citep[e.g.,][]{mueller24-outbursts}.

  \section{Conclusions} \label{sec:summary}

  We presented a 251~day lightcurve of comet 243P/NEAT, including broad-band and narrow-band color photometry.  After perihelion, the comet moved through opposition, and gained about --0.7~mag in intrinsic brightness, relative to the pre-opposition trend.  Based on a dust dynamical model of the lightcurve and comet morphology, we show that the brightening may be caused by coma and/or tail projection effects, and that the perihelion dust mass loss rate was about 2.7~\kgps.

  The lightcurve also displayed 2 to 3 outbursts, one of which was approximately --3~mag in relative brightness \citep{heinze18-cbet4587}.  We modeled an image of the outburst ejecta and the outburst lightcurve with a dust dynamical model and estimated the total ejected mass to be $(3\pm1)\times10^7$~kg.  By extension, the smaller outbursts ($\Delta m\sim -0.3$~mag) released $\sim10^5$~kg of dust.

  The large outburst of 243P likely excavated material previously buried to $\sim$10~m depths.  We reviewed the sub-surface properties of comet 9P/Tempel 1 and 67P/Churyumov-Gerasimenko.  Both of these cometary nuclei have water ice within centimeters or meters of the surface.  Therefore, by analogy, the large outburst of comet 243P likely liberated water ice grains from its nucleus.

  A near-infrared spectrum of the ejecta $4\pm1$~days after the outburst lacked evidence for water ice absorption features, with an upper-limit 2-\micron{} band depth of 3\%.  The pre-outburst and post-outburst optical colors were red and consistent with each other.  A map of the optical continuum color taken $5\pm1$~days after the outburst lacked any evidence for color gradients out to 40,000~km from the nucleus.

  Observations of water ice grains in cometary comae tend to show evidence for sub-micrometer and micrometer grain sizes.  We showed that the expansion speeds of sub-micrometer and micrometer-sized water ice grains were high enough to leave the aperture of our spectroscopic observation and preclude their detection.  However, such grain sizes would still be in the vicinity of the nucleus and observable in our optical color map taken 5~days after the outburst.  To explain the lack of a color gradient in the optical data, we showed that water ice grains in contact with a small amount of low albedo material (dust) would have sublimated away before our spectroscopic and color observations.

  For additional context on the 243P outburst observations, we analyzed near-infrared spectra and optical colors of comet 103P/Hartley 2 taken with the Deep Impact flyby spacecraft \citep{ahearn05,li13-hartley2,protopapa14}.  We found that much of the coma has a strong correlation between water ice content and optical color, with bluer regions having a higher areal fraction of ice.  However, this is not true for all observed regions in the coma.  A localized source (labeled $J_4$ by \citealt{protopapa14}) was shown to have a substantial amount of water ice, but an optical color as red as any ice-free region in the coma.  It is evident that the scattering properties of ejected dust vary in 103P, and that they are just as important to optical color as percent-level abundances of water ice.

  Given these results, we find that assessing the presence of water ice on the basis of continuum colors alone is limited, and the addition of other supporting data, such as near-infrared spectroscopy, is critical to their interpretation.  Further studies of continuum color and near-infrared spectroscopy are warranted to determine the range of dust and ice properties in the coma, and under what circumstances may we use continuum color as a proxy for water ice content.

  Finally, we also examined the specific kinetic energy of comet 243P's large outburst, $\sim10-100$~J~kg$^{-1}$, finding it to be close to the specific energy of the comet's quiescent sublimation, $\sim1-10$~J~kg$^{-1}$.  Overall, a range of specific energies has been previously found for cometary outbursts.  Based on the studies of 5 comets (including our 243P work), there appears to be a grouping in the outburst ejecta specific energies, with two comets near $~10^1-10^2$~J~\inv{kg} and three comets near $\sim10^4-10^5$~J~\inv{kg}.  More characterizations of outbursts are needed to determine if this grouping holds, or if they are instead samples of a continuum of values or some other distribution.

  The crystallization of amorphous water ice is commonly adopted to explain these events, but its presence near the surfaces of Jupiter-family comets is difficult to ascertain \citep{gudipati15}.  The specific energy of this transition is $\sim10^5$~J~kg$^{-1}$, orders of magnitude higher than what is needed to explain 243P's large outburst.  Trapped volatile gases provide a key role driving the mass-loss following water ice crystallization \citep{prialnik24-comets3}.  We propose that if the crystallization of water ice was responsible for the outburst then the cause for the difference in specific energies in, e.g., the 243P and 17P outbursts may be due to the relative abundance of trapped volatile gases.

  \bigskip
  The authors thank K.~Meech who initially shared the 243P/NEAT outburst discovery with the Comet 46P/Wirtanen Campaign email list; M.~Graham and E.~Bellm who kindly donated ARC telescope time; M.~Knight and C.~Holt who assisted with the LDT observations; C.-S.~Lin and M.~Zhang who assisted with the Lulin Observatory data; A.~Gibbs, F.~Shelly, and the Catalina Sky Survey project for supplying the 703 and G96 data and answering questions about the data; and, J.Y.~Li for sharing registered Deep Impact optical data.

  M.S.P.K. and S.P. acknowledge funding from NASA (USA) Solar System Observations program grants NNX15AD99G and 80NSSC20K0673. Observations made by L.F. and T.F. were funded by NSF grant 1852589. Q.Y. was supported by the GROWTH project funded by the National Science Foundation (USA) Grant No.\ 1545949. C.-H.H. thanks supports from the Hong Kong Research Grants Council for GRF research support under the grants 17326116 and 17300417, and thanks Q.A. Parker and HKU for provision of his research post.  B.T.B. was supported by an appointment to the NASA Postdoctoral Program at the NASA Goddard Space Flight Center, administered by Oak Ridge Associated Universities under contract with NASA.

  Based on observations obtained with the Samuel Oschin Telescope 48-inch Telescope at the Palomar Observatory as part of the Zwicky Transient Facility project. ZTF is supported by the National Science Foundation (USA) under Grant No.\ AST-1440341 and a collaboration including Caltech, IPAC, the Weizmann Institute for Science, the Oskar Klein Center at Stockholm University, the University of Maryland, the University of Washington, Deutsches Elektronen-Synchrotron and Humboldt University, Los Alamos National Laboratories, the TANGO Consortium of Taiwan, the University of Wisconsin at Milwaukee, and Lawrence Berkeley National Laboratories. Operations are conducted by COO, IPAC, and UW.

  This publication has made use of data collected at Lulin Observatory, partially supported by MoST grant 105-2112-M-008-024-MY3.

  This work makes use of observations from the LCOGT network operated by Las Cumbres Observatory.

  The Catalina Sky Survey (CSS) is funded by a NASA Near Earth Object Observations (NEOO) grant 80NSSC18K1130 to the University of Arizona.

  These results made use of the Lowell Discovery Telescope at Lowell Observatory. Lowell is a private, non-profit institution dedicated to astrophysical research and public appreciation of astronomy and operates the LDT in partnership with Boston University, the University of Maryland, the University of Toledo, Northern Arizona University and Yale University.  The Large Monolithic Imager was built by Lowell Observatory using funds provided by the National Science Foundation (AST-1005313).

  TRAPPIST-North is a project funded by the University of Liege, in collaboration with the Cadi Ayyad University of Marrakech (Morocco). TRAPPIST-South is a project funded by the Belgian Fonds (National) de la Recherche Scientifique (F.R.S.-FNRS) under grant PDR T.0120.21. E.J is an F.R.S.-FNRS Senior Research Associate.

  The operation of Xingming Observatory and the NEXT telescope was made possible by the generous support from the Xinjiang Astronomical Observatory and the Ningbo Bureau of Education.

  This research has made use of the NASA/IPAC Infrared Science Archive, which is operated by the Jet Propulsion Laboratory, California Institute of Technology, under contract with the National Aeronautics and Space Administration.

  Our work is partially based on observations obtained with the Apache Point Observatory 3.5-meter telescope, which is owned and operated by the Astrophysical Research Consortium. We thank the Director (Nancy Chanover) and Deputy Director (Ben Williams) of the Astrophysical Research Consortium (ARC) 3.5m telescope at Apache Point Observatory for their enthusiastic and timely support of our Director's Discretionary Time (DDT) proposals. We also thank Russet McMillan and the rest of the APO technical staff for their assistance in performing the observations just two days after our DDT proposals were submitted. BTB would like to acknowledge the generous support of the B612 Foundation and its Asteroid Institute program. BTB wishes to acknowledge the support of DIRAC (Data Intensive Research in Astronomy and Cosmology) Institute at the University of Washington.

  Funding for the Asteroid Institute program is provided by B612 Foundation, W.K. Bowes Jr. Foundation, P. Rawls Family Fund and two anonymous donors in addition to general support from the B612 Founding Circle (K. Algeri-Wong, B. Anders, G. Baehr, B. Burton, A. Carlson, D. Carlson, S. Cerf, V. Cerf, Y. Chapman, J. Chervenak, D. Corrigan, E. Corrigan, A. Denton, E. Dyson, A. Eustace, S. Galitsky, The Gillikin Family, E. Gillum, L. Girand, Glaser Progress Foundation, D. Glasgow, J. Grimm, S. Grimm, G. Gruener, V. K. Hsu $\&$ Sons Foundation Ltd., J. Huang, J. D. Jameson, J. Jameson, M. Jonsson Family Foundation, S. Jurvetson, D. Kaiser, S. Krausz, V. La\v{v}as, J. Leszczenski, D. Liddle, S. Mak, G.McAdoo, S. McGregor, J. Mercer, M. Mullenweg, D. Murphy, P. Norvig, S. Pishevar, R. Quindlen, N. Ramsey, R. Rothrock, E. Sahakian, R. Schweickart, A. Slater, T. Trueman, F. B. Vaughn, R. C. Vaughn, B. Wheeler, Y. Wong, M. Wyndowe, plus six anonymous donors).

  This research made use of the Horizons online ephemeris system developed and operated by the Solar System Dynamics Group at NASA's Jet Propulsion Laboratory.

  \vspace{5mm}
  \facilities{PO:1.2m, IRSA, Hale, IRTF, ATLAS, LCOGT, TRAPPIST, XMO:NEXT, LO:1m, LDT, ARC, SO:Schmidt, SO:1.5m}


  \software{
    acronym \citep{weisenburger17-acronym},
    astroimagej \citep{collins17-astroimagej}
    astropy \citep{astropy13,astropy18,astropy22},
    calviacat \citep{kelley19-calviacat},
    dct-redux \citep{kelley24-dct-redux-v0.8.5},
    IRAF \citep{tody86},
    photometrypipeline \citep{mommert2017ascl},
    pycometsuite \citep{kelley23-pycometsuite-v1.0.0},
    pyfits \citep{barret12-pyfits},
    sbpy \citep{mommert19-sbpy},
    scipy \citep{virtanen20-scipy},
    SEP \citep{barbary2018ascl},
    SExtractor \citep{bertin96},
    synphot \citep{synphot2018ascl}
  }

  \appendix
  \restartappendixnumbering

  \section{Optical observational data details}\label{app:obs}

  \begin{longrotatetable}
    \begin{deluxetable*}{ll DD ccD c DDD c DD DD}
      \tablecaption{243P/NEAT Observing Circumstances, Photometry, \afrho[\phi] Values, and Geometric Cross Sectional Area \label{tab:obs}}
      \tabletypesize{\scriptsize}
      \tablehead{
        \colhead{Source}
        & \colhead{Date}
        & \multicolumn2c{$T-T_P$}
        & \multicolumn2c{$T-T_0$}
        & \colhead{\rh}
        & \colhead{$\Delta$}
        & \multicolumn2c{$\phi$}
        & \colhead{Filters}
        & \multicolumn2c{Airmass}
        & \multicolumn2c{IQ}
        & \multicolumn2c{$\rho$}
        & \colhead{Cat. Filt.}
        & \multicolumn2c{$m$}
        & \multicolumn2c{$\sigma_m$}
        & \multicolumn2c{\afrho[\theta]}
        & \multicolumn2c{$G$}
        \\
        & \colhead{(UT)}
        & \multicolumn2c{(days)}
        & \multicolumn2c{(days)}
        & \colhead{(au)}
        & \colhead{(au)}
        & \multicolumn2c{(\degr)}
        & \colhead{}
        & \multicolumn2c{(\arcsec)}
        & \multicolumn2c{(\arcsec)}
        & \multicolumn2c{(\arcsec)}
        & \colhead{}
        & \multicolumn2c{(mag)}
        & \multicolumn2c{(mag)}
        & \multicolumn2c{(cm)}
        & \multicolumn2c{(km$^2$)}
      }
      \colnumbers
      \decimals
      \startdata
      ZTF & 2018-06-20 11:09 & -66.54 & -173.9 & 2.493 & 2.57 & 23.09 & $r$ & 2.3 & 2.8 & 5.37 & $r$ & 19.56 & 0.16 & 9.32 & 35.8 \\
      ZTF & 2018-07-13 10:39 & -43.56 & -150.92 & 2.471 & 2.288 & 24.28 & $r$ & 1.8 & 1.6 & 6.03 & $r$ & 19.41 & 0.16 & 8.33 & 32.9 \\
      ZTF & 2018-07-16 11:35 & -40.52 & -147.88 & 2.469 & 2.251 & 24.31 & $r$ & 1.4 & 1.9 & 6.13 & $r$ & 19.23 & 0.15 & 9.51 & 37.6 \\
      ZTF & 2018-07-22 10:39 & -34.56 & -141.92 & 2.465 & 2.179 & 24.28 & $g$ & 1.5 & 2.2 & 6.33 & $g$ & 19.84 & 0.15 & 7.3 & 34.0 \\
      ZTF & 2018-07-22 11:28 & -34.53 & -141.88 & 2.465 & 2.178 & 24.28 & $r$ & 1.3 & 2.2 & 6.33 & $r$ & 19.4 & 0.13 & 7.56 & 29.8 \\
      ZTF & 2018-07-25 11:36 & -31.52 & -138.88 & 2.463 & 2.142 & 24.22 & $r$ & 1.3 & 1.9 & 6.44 & $r$ & 19.19 & 0.11 & 8.85 & 34.9 \\
      ATLAS HKO & 2018-07-26 14:35 & -30.4 & -137.75 & 2.463 & 2.129 & 24.18 & $o$ & 1.2 & 3.7 & 6.48 & $o$ & 18.9 & 0.22 & 10.88 & 42.0 \\
      ZTF & 2018-08-06 11:22 & -19.53 & -126.89 & 2.458 & 2.002 & 23.57 & $r$ & 1.3 & 1.9 & 6.89 & $r$ & 19.03 & 0.11 & 8.92 & 34.7 \\
      ZTF & 2018-08-06 11:38 & -19.52 & -126.88 & 2.458 & 2.002 & 23.57 & $g$ & 1.2 & 2.6 & 6.89 & $g$ & 19.47 & 0.16 & 8.62 & 39.5 \\
      ZTF & 2018-08-19 10:59 & -6.55 & -113.9 & 2.455 & 1.86 & 22.09 & $g$ & 1.2 & 2.1 & 7.42 & $g$ & 19.39 & 0.11 & 8.01 & 35.5 \\
      ZTF & 2018-08-23 09:18 & -2.62 & -109.97 & 2.455 & 1.819 & 21.46 & $r$ & 1.5 & 2.0 & 7.58 & $r$ & 18.71 & 0.08 & 9.85 & 36.4 \\
      ZTF & 2018-08-25 10:58 & -0.55 & -107.9 & 2.454 & 1.798 & 21.09 & $r$ & 1.2 & 2.1 & 7.67 & $r$ & 18.64 & 0.09 & 10.28 & 37.7 \\
      ZTF & 2018-08-26 09:11 & 0.38 & -106.98 & 2.454 & 1.789 & 20.91 & $r$ & 1.5 & 2.3 & 7.71 & $r$ & 18.66 & 0.17 & 10.03 & 36.6 \\
      ZTF & 2018-08-27 09:30 & 1.39 & -105.97 & 2.454 & 1.779 & 20.72 & $r$ & 1.4 & 1.8 & 7.75 & $r$ & 18.57 & 0.15 & 10.78 & 39.1 \\
      ZTF & 2018-09-03 11:25 & 8.47 & -98.88 & 2.455 & 1.713 & 19.19 & $g$ & 1.2 & 2.7 & 8.05 & $g$ & 19.04 & 0.12 & 9.43 & 38.9 \\
      ATLAS MLO & 2018-09-03 14:55 & 8.62 & -98.74 & 2.455 & 1.712 & 19.15 & $o$ & 1.1 & 5.8 & 8.05 & $o$ & 18.19 & 0.09 & 13.48 & 46.1 \\
      ZTF & 2018-09-08 10:29 & 13.43 & -93.92 & 2.456 & 1.671 & 17.93 & $r$ & 1.1 & 2.1 & 8.25 & $r$ & 18.47 & 0.07 & 10.39 & 35.1 \\
      ZTF & 2018-09-08 10:52 & 13.45 & -93.91 & 2.456 & 1.671 & 17.92 & $g$ & 1.1 & 2.4 & 8.25 & $g$ & 18.91 & 0.08 & 10.08 & 40.1 \\
      ATLAS HKO & 2018-09-09 14:47 & 14.61 & -92.74 & 2.456 & 1.661 & 17.61 & $o$ & 1.2 & 3.8 & 8.3 & $o$ & 18.04 & 0.07 & 14.61 & 48.0 \\
      ZTF & 2018-09-11 09:23 & 16.38 & -90.97 & 2.457 & 1.647 & 17.11 & $r$ & 1.2 & 2.3 & 8.37 & $r$ & 18.04 & 0.05 & 15.07 & 49.8 \\
      ZTF & 2018-09-11 10:39 & 16.44 & -90.92 & 2.457 & 1.647 & 17.09 & $g$ & 1.1 & 2.7 & 8.37 & $g$ & 18.55 & 0.06 & 13.66 & 53.2 \\
      ATLAS MLO & 2018-09-11 14:32 & 16.6 & -90.75 & 2.457 & 1.645 & 17.04 & $o$ & 1.1 & 6.2 & 8.38 & $o$ & 17.92 & 0.05 & 16.06 & 52.0 \\
      ZTF & 2018-09-14 09:54 & 19.41 & -87.95 & 2.458 & 1.624 & 16.21 & $r$ & 1.2 & 2.2 & 8.49 & $r$ & 17.91 & 0.05 & 16.57 & 53.4 \\
      ZTF & 2018-09-14 10:24 & 19.43 & -87.93 & 2.458 & 1.624 & 16.2 & $g$ & 1.1 & 2.4 & 8.49 & $g$ & 18.58 & 0.07 & 12.97 & 49.3 \\
      ATLAS MLO & 2018-09-15 14:05 & 20.58 & -86.77 & 2.458 & 1.616 & 15.85 & $o$ & 1.1 & 5.8 & 8.53 & $o$ & 17.9 & 0.04 & 15.73 & 49.2 \\
      ZTF & 2018-09-17 09:36 & 22.39 & -84.96 & 2.459 & 1.603 & 15.28 & $r$ & 1.2 & 2.0 & 8.6 & $r$ & 17.99 & 0.05 & 15.02 & 47.1 \\
      ZTF & 2018-09-17 10:48 & 22.44 & -84.91 & 2.459 & 1.603 & 15.26 & $g$ & 1.2 & 2.6 & 8.6 & $g$ & 18.54 & 0.07 & 13.11 & 48.5 \\
      ATLAS HKO & 2018-09-17 13:56 & 22.57 & -84.78 & 2.459 & 1.602 & 15.22 & $o$ & 1.1 & 3.9 & 8.61 & $o$ & 17.86 & 0.06 & 16.1 & 49.4 \\
      ATLAS MLO & 2018-09-19 13:52 & 24.57 & -82.78 & 2.46 & 1.589 & 14.56 & $o$ & 1.1 & 6.8 & 8.68 & $o$ & 17.93 & 0.06 & 14.8 & 44.6 \\
      ZTF & 2018-09-21 08:36 & 26.35 & -81.0 & 2.461 & 1.578 & 13.96 & $r$ & 1.2 & 2.2 & 8.74 & $r$ & 17.88 & 0.07 & 16.14 & 48.7 \\
      ZTF & 2018-09-21 10:23 & 26.43 & -80.93 & 2.461 & 1.578 & 13.94 & $g$ & 1.1 & 2.1 & 8.74 & $g$ & 18.5 & 0.07 & 13.11 & 46.6 \\
      ATLAS HKO & 2018-09-21 14:00 & 26.58 & -80.78 & 2.461 & 1.577 & 13.89 & $o$ & 1.1 & 4.1 & 8.74 & $o$ & 17.79 & 0.06 & 16.67 & 49.2 \\
      ZTF & 2018-09-22 08:32 & 27.35 & -80.0 & 2.461 & 1.572 & 13.62 & $r$ & 1.3 & 1.7 & 8.77 & $r$ & 18.06 & 0.09 & 13.55 & 40.5 \\
      ZTF & 2018-09-24 07:58 & 29.33 & -78.03 & 2.462 & 1.561 & 12.93 & $r$ & 1.3 & 1.7 & 8.84 & $r$ & 17.68 & 0.11 & 18.92 & 55.4 \\
      ZTF & 2018-09-24 09:24 & 29.38 & -77.97 & 2.462 & 1.561 & 12.91 & $g$ & 1.1 & 3.8 & 8.84 & $g$ & 18.2 & 0.18 & 16.98 & 58.5 \\
      ZTF & 2018-09-29 06:12 & 34.25 & -73.1 & 2.465 & 1.537 & 11.14 & $r$ & 1.7 & 2.0 & 8.97 & $r$ & 17.69 & 0.09 & 18.26 & 50.4 \\
      ZTF & 2018-09-29 09:07 & 34.37 & -72.98 & 2.465 & 1.536 & 11.1 & $g$ & 1.2 & 1.9 & 8.98 & $g$ & 18.39 & 0.1 & 13.83 & 45.0 \\
      ATLAS MLO & 2018-10-01 13:30 & 36.56 & -70.8 & 2.466 & 1.527 & 10.28 & $o$ & 1.2 & 5.4 & 9.03 & $o$ & 17.76 & 0.06 & 16.15 & 42.5 \\
      ATLAS HKO & 2018-10-03 13:06 & 38.54 & -68.81 & 2.467 & 1.52 & 9.54 & $o$ & 1.1 & 4.1 & 9.07 & $o$ & 17.62 & 0.06 & 18.1 & 46.5 \\
      ZTF & 2018-10-06 09:00 & 41.37 & -65.99 & 2.469 & 1.511 & 8.48 & $g$ & 1.1 & 2.1 & 9.13 & $g$ & 18.18 & 0.05 & 16.39 & 48.8 \\
      ZTF & 2018-10-06 09:37 & 41.39 & -65.96 & 2.469 & 1.511 & 8.47 & $r$ & 1.2 & 1.6 & 9.13 & $r$ & 17.64 & 0.04 & 18.46 & 46.6 \\
      ATLAS HKO & 2018-10-11 13:06 & 46.54 & -60.82 & 2.473 & 1.5 & 6.62 & $c$ & 1.2 & 4.6 & 9.19 & $c$ & 17.97 & 0.07 & 16.55 & 34.2 \\
      ATLAS HKO & 2018-10-15 12:12 & 50.5 & -56.85 & 2.477 & 1.496 & 5.4 & $o$ & 1.1 & 4.0 & 9.21 & $o$ & 17.31 & 0.04 & 23.67 & 52.5 \\
      ZTF & 2018-11-01 07:29 & 67.31 & -40.05 & 2.494 & 1.524 & 6.24 & $r$ & 1.3 & 2.0 & 9.05 & $r$ & 17.33 & 0.04 & 25.53 & 59.5 \\
      ZTF & 2018-11-02 06:54 & 68.28 & -39.07 & 2.495 & 1.528 & 6.56 & $r$ & 1.2 & 2.5 & 9.03 & $r$ & 17.31 & 0.12 & 26.25 & 61.9 \\
      ATLAS MLO & 2018-11-02 11:19 & 68.47 & -38.89 & 2.495 & 1.529 & 6.62 & $o$ & 1.2 & 5.7 & 9.02 & $o$ & 17.23 & 0.04 & 26.9 & 62.4 \\
      ZTF & 2018-11-03 06:44 & 69.27 & -38.08 & 2.496 & 1.532 & 6.89 & $r$ & 1.2 & 1.9 & 9.0 & $r$ & 17.29 & 0.03 & 26.97 & 64.4 \\
      ZTF & 2018-11-05 07:56 & 71.32 & -36.03 & 2.499 & 1.542 & 7.59 & $g$ & 1.3 & 3.9 & 8.94 & $g$ & 17.97 & 0.07 & 21.19 & 61.1 \\
      ATLAS MLO & 2018-11-06 11:00 & 72.45 & -34.9 & 2.5 & 1.547 & 7.99 & $o$ & 1.2 & 5.6 & 8.91 & $o$ & 17.25 & 0.03 & 27.06 & 65.9 \\
      ZTF & 2018-11-07 05:38 & 73.23 & -34.13 & 2.501 & 1.551 & 8.27 & $g$ & 1.3 & 2.4 & 8.89 & $g$ & 17.92 & 0.05 & 22.43 & 66.3 \\
      ZTF & 2018-11-07 06:31 & 73.27 & -34.09 & 2.501 & 1.552 & 8.28 & $r$ & 1.3 & 1.9 & 8.89 & $r$ & 17.34 & 0.04 & 26.39 & 66.2 \\
      ZTF & 2018-11-08 07:07 & 74.29 & -33.06 & 2.502 & 1.557 & 8.65 & $g$ & 1.2 & 2.2 & 8.86 & $g$ & 17.99 & 0.05 & 21.26 & 63.7 \\
      ATLAS HKO & 2018-11-08 11:22 & 74.47 & -32.89 & 2.502 & 1.558 & 8.71 & $c$ & 1.3 & 4.6 & 8.85 & $c$ & 17.82 & 0.05 & 21.02 & 46.8 \\
      ZTF & 2018-11-09 08:24 & 75.34 & -32.01 & 2.504 & 1.563 & 9.02 & $r$ & 1.4 & 5.0 & 8.82 & $r$ & 17.54 & 0.07 & 22.38 & 57.6 \\
      ZTF & 2018-11-10 05:45 & 76.23 & -31.12 & 2.505 & 1.568 & 9.34 & $g$ & 1.3 & 2.6 & 8.8 & $g$ & 17.95 & 0.06 & 22.3 & 68.4 \\
      ZTF & 2018-11-10 06:34 & 76.27 & -31.09 & 2.505 & 1.569 & 9.36 & $r$ & 1.3 & 2.5 & 8.79 & $r$ & 17.36 & 0.04 & 26.64 & 69.3 \\
      ATLAS MLO & 2018-11-10 10:50 & 76.45 & -30.91 & 2.505 & 1.57 & 9.42 & $o$ & 1.2 & 5.9 & 8.78 & $o$ & 17.37 & 0.04 & 25.09 & 64.2 \\
      ATLAS HKO & 2018-11-12 10:56 & 78.45 & -28.91 & 2.508 & 1.582 & 10.14 & $c$ & 1.3 & 4.8 & 8.71 & $c$ & 17.84 & 0.04 & 21.35 & 49.9 \\
      ZTF & 2018-11-13 06:31 & 79.27 & -28.09 & 2.509 & 1.588 & 10.43 & $r$ & 1.3 & 4.9 & 8.69 & $r$ & 17.45 & 0.06 & 25.03 & 67.6 \\
      ZTF & 2018-11-14 04:46 & 80.19 & -27.16 & 2.51 & 1.594 & 10.75 & $g$ & 1.3 & 5.0 & 8.65 & $g$ & 18.13 & 0.07 & 19.63 & 63.1 \\
      ZTF & 2018-11-14 07:16 & 80.3 & -27.06 & 2.51 & 1.595 & 10.79 & $r$ & 1.3 & 3.3 & 8.65 & $r$ & 17.51 & 0.09 & 23.95 & 65.4 \\
      ATLAS MLO & 2018-11-14 10:22 & 80.43 & -26.93 & 2.51 & 1.596 & 10.83 & $o$ & 1.2 & 5.7 & 8.64 & $o$ & 17.44 & 0.04 & 24.55 & 65.8 \\
      ZTF & 2018-11-15 05:32 & 81.22 & -26.13 & 2.511 & 1.602 & 11.11 & $g$ & 1.3 & 3.7 & 8.61 & $g$ & 18.16 & 0.08 & 19.3 & 62.8 \\
      ZTF & 2018-11-15 06:27 & 81.26 & -26.09 & 2.511 & 1.602 & 11.13 & $r$ & 1.3 & 2.5 & 8.61 & $r$ & 17.55 & 0.05 & 23.27 & 64.2 \\
      ZTF & 2018-11-16 06:31 & 82.27 & -25.09 & 2.513 & 1.609 & 11.47 & $r$ & 1.3 & 2.8 & 8.57 & $r$ & 17.55 & 0.14 & 23.6 & 65.9 \\
      ZTF & 2018-11-17 07:23 & 83.3 & -24.05 & 2.514 & 1.617 & 11.83 & $r$ & 1.3 & 2.1 & 8.53 & $r$ & 17.59 & 0.07 & 22.99 & 64.9 \\
      ZTF & 2018-11-25 08:05 & 91.33 & -16.02 & 2.526 & 1.685 & 14.41 & $r$ & 1.5 & 2.9 & 8.19 & $r$ & 17.73 & 0.09 & 22.14 & 67.8 \\
      ATLAS MLO & 2018-11-26 09:43 & 92.4 & -14.96 & 2.528 & 1.696 & 14.72 & $o$ & 1.2 & 5.9 & 8.13 & $o$ & 17.69 & 0.05 & 22.15 & 67.0 \\
      ATLAS HKO & 2018-11-28 09:50 & 94.4 & -12.95 & 2.531 & 1.715 & 15.3 & $o$ & 1.3 & 4.7 & 8.04 & $o$ & 17.81 & 0.09 & 20.49 & 63.1 \\
      ATLAS MLO & 2018-11-30 09:31 & 96.39 & -10.96 & 2.534 & 1.735 & 15.85 & $o$ & 1.2 & 5.6 & 7.95 & $o$ & 17.84 & 0.04 & 20.36 & 63.7 \\
      ZTF & 2018-12-04 05:25 & 100.22 & -7.13 & 2.54 & 1.776 & 16.84 & $r$ & 1.2 & 2.9 & 7.77 & $r$ & 17.92 & 0.04 & 20.91 & 68.6 \\
      ATLAS MLO & 2018-12-04 08:42 & 100.36 & -7.0 & 2.54 & 1.777 & 16.88 & $o$ & 1.2 & 6.3 & 7.76 & $o$ & 17.91 & 0.06 & 20.09 & 64.7 \\
      ZTF & 2018-12-05 05:12 & 101.21 & -6.14 & 2.542 & 1.786 & 17.08 & $r$ & 1.2 & 3.4 & 7.72 & $r$ & 17.99 & 0.04 & 19.93 & 65.8 \\
      ATLAS HKO & 2018-12-06 08:35 & 102.35 & -5.0 & 2.544 & 1.799 & 17.35 & $c$ & 1.2 & 4.9 & 7.66 & $c$ & 18.49 & 0.06 & 15.59 & 45.4 \\
      ATLAS MLO & 2018-12-08 08:29 & 104.35 & -3.01 & 2.547 & 1.822 & 17.8 & $o$ & 1.1 & 5.6 & 7.57 & $o$ & 17.96 & 0.06 & 20.34 & 67.2 \\
      CSS (703) & 2018-12-10 04:26 & 106.18 & -1.18 & 2.55 & 1.843 & 18.2 & none & \nodata & \nodata & 7.5 & $r$ & 18.21 & 0.05 & 17.34 & 59.2 \\
      ATLAS HKO & 2018-12-10 08:48 & 106.36 & -0.99 & 2.551 & 1.845 & 18.23 & $c$ & 1.2 & 5.1 & 7.47 & $c$ & 18.49 & 0.06 & 16.56 & 49.3 \\
      ATLAS MLO & 2018-12-12 08:06 & 108.33 & 0.98 & 2.554 & 1.869 & 18.63 & $o$ & 1.1 & 6.0 & 7.38 & $o$ & 15.40 & 0.01 & 227.52 & 768.4 \\
      CSS (703) & 2018-12-14 03:17 & 110.13 & 2.78 & 2.557 & 1.891 & 18.97 & none & \nodata & \nodata & 7.3 & $r$ & 15.90 & 0.06 & 154.26 & 536.5 \\
      CSS (G96) & 2018-12-14 03:50 & 110.15 & 2.8 & 2.557 & 1.891 & 18.97 & none & \nodata & \nodata & 7.3 & $r$ & 15.82 & 0.07 & 166.08 & 577.8 \\
      ZTF & 2018-12-14 05:55 & 110.24 & 2.89 & 2.557 & 1.892 & 18.99 & $r$ & 1.3 & 2.1 & 7.29 & $r$ & 15.90 & 0.04 & 153.89 & 535.2 \\
      LCOGT (T04) & 2018-12-14 08:06 & 110.33 & 2.98 & 2.558 & 1.894 & 19.01 & none & 1.1 & 3.9 & 7.3 & $r$ & 16.02 & 0.07 & 138.34 & 482.2 \\
      LCOGT (V37) & 2018-12-15 01:00 & 111.04 & 3.68 & 2.559 & 1.902 & 19.13 & $g',r'$ & 1.3 & 2.4 & 7.3 & $r$ & 16.27 & 0.04 & 110.5 & 388.3 \\
      ARC & 2018-12-15 08:03 & 111.33 & 3.98 & 2.559 & 1.906 & 19.18 & $R$ & 3.8 & 2.4 & 7.3 & $r$ & 16.38 & 0.05 & 100.09 & 352.8 \\
      NEXT & 2018-12-15 17:52 & 111.74 & 4.38 & 2.56 & 1.911 & 19.26 & $V,R$ & \nodata & \nodata & 7.2 & $r$ & 16.61 & 0.03 & 82.38 & 287.7 \\
      LCOGT (K93) & 2018-12-15 19:06 & 111.79 & 4.43 & 2.56 & 1.912 & 19.26 & $g',r'$ & 1.2 & 2.6 & 7.2 & $r$ & 16.5 & 0.03 & 91.2 & 318.7 \\
      TRAPPIST-N & 2018-12-15 21:34 & 111.89 & 4.54 & 2.56 & 1.913 & 19.28 & $B,V,R,I$ & 1.2 & 1.8 & 6.3 & $r$ & 16.65 & 0.1 & 90.85 & 278.1 \\
      LCOGT (W87) & 2018-12-16 01:06 & 112.04 & 4.68 & 2.561 & 1.915 & 19.31 & $g',r'$ & 1.2 & 2.5 & 7.2 & $r$ & 16.6 & 0.03 & 83.34 & 292.0 \\
      LDT & 2018-12-16 02:05 & 112.08 & 4.73 & 2.561 & 1.915 & 19.31 & $BC,RC,r'$ & 1.3 & 0.8 & 7.2 & $r$ & 16.54 & 0.02 & 88.1 & 308.9 \\
      Hale & 2018-12-16 05:31 & 112.22 & 4.87 & 2.561 & 1.917 & 19.34 & $r'$ & \nodata & \nodata & 7.2 & $r$ & 16.53 & 0.03 & 89.02 & 312.6 \\
      ATLAS MLO & 2018-12-16 07:20 & 112.3 & 4.95 & 2.561 & 1.918 & 19.35 & $o$ & 1.1 & 5.7 & 7.19 & $o$ & 16.5 & 0.02 & 87.01 & 299.4 \\
      LCOGT (Q63) & 2018-12-16 10:09 & 112.42 & 5.06 & 2.561 & 1.92 & 19.37 & $g',r'$ & 1.2 & 2.2 & 7.2 & $r$ & 16.62 & 0.05 & 82.06 & 288.8 \\
      Lulin & 2018-12-19 14:03 & 115.58 & 8.22 & 2.567 & 1.96 & 19.88 & $V,R$ & 1.3 & 1.4 & 7.0 & $r$ & 17.09 & 0.14 & 56.16 & 198.7 \\
      ZTF & 2018-12-20 02:52 & 116.11 & 8.76 & 2.568 & 1.967 & 19.95 & $r$ & 1.2 & 1.8 & 7.01 & $r$ & 17.17 & 0.05 & 52.52 & 187.1 \\
      ZTF & 2018-12-20 04:03 & 116.16 & 8.81 & 2.568 & 1.968 & 19.96 & $g$ & 1.2 & 2.1 & 7.01 & $g$ & 17.66 & 0.08 & 48.22 & 202.7 \\
      ATLAS MLO & 2018-12-20 08:03 & 116.33 & 8.98 & 2.569 & 1.97 & 19.99 & $o$ & 1.2 & 6.1 & 7.0 & $o$ & 17.25 & 0.07 & 46.44 & 162.4 \\
      Lulin & 2018-12-20 14:22 & 116.59 & 9.24 & 2.569 & 1.973 & 20.02 & $V,R$ & 1.4 & 1.3 & 7.0 & $r$ & 17.25 & 0.11 & 48.87 & 174.7 \\
      LCOGT (K91) & 2018-12-20 19:10 & 116.79 & 9.44 & 2.569 & 1.976 & 20.05 & $g',r'$ & 1.3 & 1.6 & 7.0 & $r$ & 17.29 & 0.09 & 47.17 & 169.0 \\
      LCOGT (W85) & 2018-12-22 00:48 & 118.03 & 10.67 & 2.572 & 1.992 & 20.22 & $g',r'$ & 1.2 & 2.3 & 6.9 & $r$ & 17.45 & 0.08 & 41.72 & 149.2 \\
      ATLAS HKO & 2018-12-22 09:17 & 118.38 & 11.03 & 2.573 & 1.997 & 20.27 & $o$ & 1.6 & 4.3 & 6.9 & $o$ & 17.39 & 0.12 & 42.19 & 148.6 \\
      LCOGT (V37) & 2018-12-23 01:24 & 119.05 & 11.7 & 2.574 & 2.006 & 20.36 & $g',r'$ & 1.2 & 2.2 & 6.9 & $r$ & 17.49 & 0.07 & 40.54 & 146.5 \\
      ZTF & 2018-12-23 02:19 & 119.09 & 11.74 & 2.574 & 2.007 & 20.36 & $g$ & 1.2 & 2.1 & 6.87 & $g$ & 18.03 & 0.13 & 35.84 & 152.1 \\
      ZTF & 2018-12-23 03:33 & 119.14 & 11.79 & 2.574 & 2.007 & 20.37 & $r$ & 1.2 & 1.9 & 6.87 & $r$ & 17.46 & 0.05 & 42.01 & 151.3 \\
      TRAPPIST-N & 2018-12-23 18:41 & 119.77 & 12.42 & 2.575 & 2.016 & 20.45 & $B,V,R,I$ & 1.2 & 3.4 & 6.3 & $r$ & 17.71 & 0.1 & 36.47 & 121.2 \\
      ATLAS MLO & 2018-12-24 08:03 & 120.33 & 12.98 & 2.576 & 2.023 & 20.52 & $o$ & 1.2 & 5.4 & 6.82 & $o$ & 17.47 & 0.06 & 40.34 & 143.0 \\
      LCOGT (W85) & 2018-12-25 00:50 & 121.03 & 13.67 & 2.578 & 2.033 & 20.6 & $g',r'$ & 1.2 & 2.4 & 6.8 & $r$ & 17.69 & 0.05 & 34.78 & 126.2 \\
      TRAPPIST-N & 2018-12-25 22:36 & 121.93 & 14.58 & 2.579 & 2.045 & 20.71 & $B,V,R,I$ & 1.4 & 2.0 & 6.3 & $r$ & 17.82 & 0.1 & 33.55 & 113.8 \\
      ATLAS HKO & 2018-12-26 07:49 & 122.32 & 14.96 & 2.58 & 2.05 & 20.75 & $o$ & 1.2 & 3.9 & 6.72 & $o$ & 17.71 & 0.05 & 33.16 & 118.2 \\
      LCOGT (Q64) & 2018-12-26 10:46 & 122.44 & 15.09 & 2.58 & 2.052 & 20.76 & $g',r'$ & 1.3 & 1.8 & 6.7 & $r$ & 17.75 & 0.03 & 33.79 & 122.5 \\
      ATLAS MLO & 2018-12-28 07:40 & 124.31 & 16.96 & 2.584 & 2.078 & 20.96 & $o$ & 1.2 & 5.5 & 6.64 & $o$ & 17.75 & 0.03 & 33.01 & 118.3 \\
      TRAPPIST-N & 2018-12-29 21:40 & 125.9 & 18.54 & 2.587 & 2.1 & 21.11 & $B,V,R,I$ & 1.3 & 2.5 & 6.3 & $r$ & 17.99 & 0.1 & 29.63 & 104.2 \\
      ARC & 2018-12-31 01:20 & 127.05 & 19.69 & 2.59 & 2.116 & 21.21 & $V,R$ & 1.2 & 1.2 & 6.5 & $r$ & 17.88 & 0.05 & 32.09 & 117.6 \\
      ZTF & 2018-12-31 02:49 & 127.11 & 19.76 & 2.59 & 2.117 & 21.22 & $g$ & 1.1 & 3.2 & 6.52 & $g$ & 18.55 & 0.07 & 25.09 & 108.9 \\
      TRAPPIST-N & 2019-01-01 23:11 & 128.96 & 21.6 & 2.593 & 2.143 & 21.37 & $B,V,R,I$ & 1.8 & 2.3 & 6.3 & $r$ & 18.35 & 0.1 & 21.81 & 78.8 \\
      ZTF & 2019-01-03 02:34 & 130.1 & 22.75 & 2.596 & 2.159 & 21.45 & $r$ & 1.2 & 2.2 & 6.39 & $r$ & 18.05 & 0.05 & 28.67 & 106.0 \\
      ZTF & 2019-01-04 02:58 & 131.12 & 23.76 & 2.598 & 2.174 & 21.52 & $r$ & 1.2 & 1.7 & 6.34 & $r$ & 18.16 & 0.04 & 26.18 & 96.9 \\
      ZTF & 2019-01-04 03:37 & 131.14 & 23.79 & 2.598 & 2.174 & 21.52 & $g$ & 1.2 & 2.1 & 6.34 & $g$ & 18.75 & 0.05 & 22.09 & 96.4 \\
      ATLAS MLO & 2019-01-05 07:19 & 132.3 & 24.94 & 2.6 & 2.191 & 21.59 & $o$ & 1.2 & 5.6 & 6.29 & $o$ & 18.04 & 0.05 & 28.4 & 103.3 \\
      ATLAS HKO & 2019-01-07 07:36 & 134.31 & 26.96 & 2.605 & 2.22 & 21.69 & $c$ & 1.3 & 4.7 & 6.21 & $c$ & 18.51 & 0.05 & 24.57 & 79.9 \\
      ZTF & 2019-01-08 02:38 & 135.1 & 27.75 & 2.606 & 2.231 & 21.73 & $r$ & 1.1 & 3.0 & 6.18 & $r$ & 18.31 & 0.06 & 24.33 & 90.6 \\
      ZTF & 2019-01-08 03:51 & 135.15 & 27.8 & 2.606 & 2.232 & 21.73 & $g$ & 1.2 & 2.8 & 6.18 & $g$ & 18.77 & 0.07 & 23.05 & 101.2 \\
      ATLAS MLO & 2019-01-09 07:07 & 136.29 & 28.94 & 2.609 & 2.249 & 21.78 & $o$ & 1.2 & 5.6 & 6.13 & $o$ & 18.27 & 0.07 & 24.48 & 89.5 \\
      ZTF & 2019-01-10 02:36 & 137.1 & 29.75 & 2.61 & 2.261 & 21.81 & $r$ & 1.1 & 2.2 & 6.1 & $r$ & 18.34 & 0.04 & 24.41 & 91.1 \\
      ZTF & 2019-01-10 03:59 & 137.16 & 29.81 & 2.611 & 2.261 & 21.81 & $g$ & 1.3 & 2.7 & 6.1 & $g$ & 19.04 & 0.09 & 18.46 & 81.2 \\
      ZTF & 2019-01-11 02:38 & 138.1 & 30.75 & 2.613 & 2.275 & 21.84 & $r$ & 1.1 & 2.5 & 6.06 & $r$ & 18.37 & 0.06 & 23.95 & 89.4 \\
      ZTF & 2019-01-11 03:50 & 138.15 & 30.8 & 2.613 & 2.276 & 21.84 & $g$ & 1.2 & 2.3 & 6.06 & $g$ & 19.0 & 0.1 & 19.51 & 85.9 \\
      ATLAS HKO & 2019-01-11 06:47 & 138.28 & 30.92 & 2.613 & 2.278 & 21.85 & $c$ & 1.1 & 5.6 & 6.05 & $c$ & 18.72 & 0.08 & 21.42 & 69.9 \\
      LDT & 2019-01-12 02:13 & 139.09 & 31.73 & 2.615 & 2.29 & 21.87 & $r'$ & 1.1 & 2.1 & 6.0 & $r$ & 18.56 & 0.02 & 20.5 & 76.3 \\
      ZTF & 2019-01-12 03:53 & 139.16 & 31.8 & 2.615 & 2.291 & 21.87 & $g$ & 1.3 & 2.4 & 6.02 & $g$ & 19.11 & 0.17 & 17.87 & 78.7 \\
      ATLAS MLO & 2019-01-13 06:12 & 140.25 & 32.9 & 2.617 & 2.307 & 21.89 & $o$ & 1.1 & 5.4 & 5.98 & $o$ & 18.44 & 0.07 & 22.04 & 80.8 \\
      ATLAS MLO & 2019-01-17 06:17 & 144.26 & 36.9 & 2.626 & 2.367 & 21.94 & $o$ & 1.1 & 5.8 & 5.83 & $o$ & 18.32 & 0.15 & 26.06 & 95.6 \\
      ZTF & 2019-01-20 02:35 & 147.1 & 39.75 & 2.632 & 2.41 & 21.94 & $r$ & 1.1 & 3.1 & 5.72 & $r$ & 18.51 & 0.15 & 24.09 & 90.1 \\
      ATLAS HKO & 2019-01-20 07:01 & 147.29 & 39.93 & 2.633 & 2.412 & 21.93 & $o$ & 1.3 & 4.2 & 5.71 & $o$ & 18.49 & 0.23 & 23.37 & 85.7 \\
      ATLAS MLO & 2019-01-21 06:18 & 148.26 & 40.9 & 2.635 & 2.427 & 21.92 & $o$ & 1.2 & 5.5 & 5.68 & $o$ & 18.59 & 0.09 & 21.53 & 79.0 \\
      ATLAS HKO & 2019-01-21 06:44 & 148.27 & 40.92 & 2.635 & 2.427 & 21.92 & $o$ & 1.2 & 4.0 & 5.68 & $o$ & 18.58 & 0.15 & 21.86 & 80.2 \\
      ZTF & 2019-01-22 02:42 & 149.11 & 41.75 & 2.637 & 2.44 & 21.91 & $r$ & 1.2 & 2.9 & 5.65 & $r$ & 18.73 & 0.08 & 20.14 & 75.3 \\
      ATLAS MLO & 2019-01-22 05:22 & 149.22 & 41.86 & 2.637 & 2.441 & 21.91 & $o$ & 1.1 & 5.3 & 5.65 & $o$ & 18.86 & 0.18 & 17.0 & 62.3 \\
      ATLAS HKO & 2019-01-23 06:14 & 150.25 & 42.9 & 2.64 & 2.457 & 21.89 & $o$ & 1.1 & 4.2 & 5.61 & $o$ & 18.64 & 0.09 & 21.33 & 78.1 \\
      ZTF & 2019-01-24 03:57 & 151.16 & 43.8 & 2.642 & 2.471 & 21.87 & $g$ & 1.3 & 2.4 & 5.58 & $g$ & 19.46 & 0.1 & 15.38 & 67.7 \\
      TRAPPIST-N & 2019-01-24 19:13 & 151.79 & 44.44 & 2.643 & 2.48 & 21.86 & $B,V,R,I$ & 1.1 & 2.5 & 6.3 & $r$ & 18.8 & 0.1 & 17.33 & 73.3 \\
      ATLAS MLO & 2019-01-25 05:49 & 152.24 & 44.88 & 2.644 & 2.487 & 21.84 & $o$ & 1.1 & 5.7 & 5.54 & $o$ & 18.77 & 0.08 & 19.41 & 71.1 \\
      ZTF & 2019-01-26 04:23 & 153.18 & 45.82 & 2.646 & 2.501 & 21.82 & $g$ & 1.4 & 3.7 & 5.52 & $g$ & 19.37 & 0.15 & 17.21 & 75.8 \\
      ATLAS HKO & 2019-01-27 05:19 & 154.22 & 46.86 & 2.649 & 2.517 & 21.78 & $o$ & 1.1 & 4.6 & 5.48 & $o$ & 18.8 & 0.11 & 19.38 & 70.8 \\
      ZTF & 2019-01-28 03:27 & 155.14 & 47.78 & 2.651 & 2.531 & 21.75 & $g$ & 1.2 & 2.9 & 5.45 & $g$ & 19.39 & 0.14 & 17.39 & 76.4 \\
      ATLAS MLO & 2019-01-29 05:15 & 156.21 & 48.86 & 2.654 & 2.548 & 21.71 & $o$ & 1.1 & 6.6 & 5.41 & $o$ & 18.98 & 0.21 & 16.86 & 61.5 \\
      ZTF & 2019-01-31 03:34 & 158.14 & 50.79 & 2.658 & 2.577 & 21.62 & $g$ & 1.2 & 1.8 & 5.35 & $g$ & 19.49 & 0.12 & 16.43 & 71.9 \\
      ATLAS MLO & 2019-01-31 05:21 & 158.22 & 50.86 & 2.658 & 2.578 & 21.62 & $o$ & 1.1 & 5.6 & 5.29 & $o$ & 19.0 & 0.15 & 17.15 & 61.7 \\
      ATLAS HKO & 2019-02-04 06:44 & 162.27 & 54.92 & 2.668 & 2.64 & 21.4 & $c$ & 1.4 & 5.0 & 5.22 & $c$ & 19.2 & 0.09 & 19.36 & 62.5 \\
      TRAPPIST-S & 2019-02-05 01:13 & 163.04 & 55.69 & 2.67 & 2.651 & 21.35 & $B,V,R$ & 2.0 & 3.3 & 6.3 & $r$ & 18.97 & 0.1 & 16.16 & 72.2 \\
      ZTF & 2019-02-09 03:11 & 167.13 & 59.77 & 2.68 & 2.713 & 21.07 & $g$ & 1.3 & 2.3 & 5.08 & $g$ & 19.56 & 0.14 & 17.47 & 75.5 \\
      ZTF & 2019-02-09 04:07 & 167.17 & 59.81 & 2.68 & 2.714 & 21.06 & $r$ & 1.5 & 2.5 & 5.08 & $r$ & 18.99 & 0.11 & 20.27 & 74.2 \\
      ATLAS HKO & 2019-02-09 06:53 & 167.28 & 59.93 & 2.68 & 2.716 & 21.06 & $c$ & 1.5 & 5.3 & 5.08 & $c$ & 19.31 & 0.15 & 18.55 & 59.4 \\
      ATLAS MLO & 2019-02-14 05:28 & 172.22 & 64.87 & 2.693 & 2.79 & 20.65 & $o$ & 1.2 & 5.6 & 4.84 & $o$ & 18.74 & 0.14 & 26.51 & 92.3 \\
      ZTF & 2019-02-24 02:35 & 182.1 & 74.75 & 2.718 & 2.937 & 19.66 & $r$ & 1.2 & 2.5 & 4.7 & $r$ & 19.32 & 0.16 & 17.99 & 63.7 \\
      TRAPPIST-S & 2019-02-26 00:31 & 184.01 & 76.66 & 2.723 & 2.966 & 19.44 & $R$ & 2.3 & 3.8 & 6.3 & $r$ & 19.12 & 0.1 & 16.38 & 78.0 \\
      \enddata
      \tablecomments{Columns: (1) telescope (IAU observatory code) ; (2) date of observation; (3) time from perihelion date, 2018 Aug 26 00:10 UT; (4) time from nominal outburst date, 2018 Dec 11 08:40 UT; (5) heliocentric distance; (6) observer-comet distance; (7) phase angle; (8) observation filters; (10) Image Quality (FWHM of stars; the LCO and Lulin data are potentially broadened by stellar trailing (Appendix~\ref{app:obs})); (11) photometric aperture radius, most are set to 10$^4$~km at the distance of the comet; (12) catalog bandpass for reported magnitude (12) apparent magnitude; (13) apparent magnitude uncertainty; (14) comet dust parameter (not corrected to 0\degr{} phase); (15) effective geometric cross sectional area.}
    \end{deluxetable*}
  \end{longrotatetable}

  The optical data and data reduction procedures were summarized in Section~\ref{sec:observations}.  The following sub-sections provide details on each observational data set, including deviations from the general procedures.

  \subsection{Zwicky Transient Facility}

  The ZTF is an optical, wide-field, time-domain survey utilizing the 48-in Samuel Oschin Telescope at Palomar Observatory \citep{dekany20-ztf-observing-sys,bellm19-ztf}.  The survey has a wide range of science programs, including the study of comets \citep{graham19-ztf}.  The camera has a 55-deg$^2$ field-of-view, with a nominal pixel scale of 1\farcs01, and a fill-factor of 87.5\% on the sky.  We searched the ZTF Archive with the ZChecker program \citep{kelley19-zchecker} for images covering the ephemeris position of comet 243P around the time of perihelion: 78 $g$- and $r$-band images with 30-s exposure times between 2018 June 20 and 2019 February 24 were found.  The data are all publicly available in the ZTF Image Service \citep{ztf-image-service}.  The images are calibrated to the Pan-STARRS~1 (PS1) photometric catalog \citep{masci19-ztf,chambers16-ps1}.  Photometry of the comet was measured within a 10,000~km radius aperture.  Due to sidereal tracking, the comet was trailed, but no more than 0.5~pix. Within a 5-pix aperture, 0.5-pix of trailing of a $1/\rho$ coma results in 0.006~mag of losses, which does not affect the results.

  \subsection{Hale Telescope}

  We also conducted follow-up observations using the 5.1-m Hale Telescope and the Wafer-Scale Imager for Prime (WaSP) camera (with a pixel size of 0\farcs18). Images were taken on 2018 December 16 UTC using the SDSS $r'$ filter, with 60~s exposures and a total integration time of 300~s. Calibration data included dome flat fields.  Color-corrected photometry was then derived using the Pan-STARRS~1 DR1 catalog \citep{chambers16-ps1,flewelling20-ps1cat}, with a 7\farcs2-radius aperture (equal to 10,000~km at the comet).  The stars were not significantly trailed.

  \subsection{Asteroid Terrestrial-Impact Last Alert System}

  ATLAS \citep{tonry18-atlas} is an optical, widefield survey system using four 0.5-m F/2 Schmidt telescopes (two in Hawaii, one in South Africa, and one in Chile) equipped with monolithic 10560$\times$10560 pixel CCD cameras delivering a 29 deg$^2$ field of view with 1.86 arcsecond pixels. On a good night, each ATLAS unit acquires 700--1000 images with 30-s exposure times, and surveys about one fourth of the sky accessible from its latitude, taking four images of each field over a time span of 30--60 minutes. ATLAS produces precisely calibrated photometric observations in two customized wideband filters, the ``cyan'' ($c$) band from 420--650 nm, and the ``orange'' ($o$) band from 560--820 nm \citep[for detailed analysis, see][]{tonry18-atlas}.  ATLAS acquired 203 photometric measurements of 243P in the time span relevant to this work (see Figure \ref{fig:lightcurve}). All measurements used a time-varying 10,000~km radius aperture. They were individually calibrated to account for aperture losses due to the small aperture-to-image quality ratio (see Table~\ref{tab:obs}), and then reducexd to a consistent magnitude scale across the different filters. Due to sidereal tracking, the comet was trailed, but no more than 0.3~pix, which did not affect the results.

  \subsection{Las Cumbres Observatory}

  Images of the comet were also taken with the Las Cumbres Observatory Global Telescope (LCOGT) network.  We scheduled a series of observations using the 1-m telescopes' Sinistro cameras (pixel size 0\farcs39, 4k$\times$4k CCDs), and SDSS $r'$ and $g'$ filters under program NOAO2019A-010.  We also include two unfiltered images of 243P taken as part of an education and public outreach program (LCOEPO2014B-010) with a 0.4-m telescope on 2018 December 14 (SBIG STX6303 camera, pixel size 0\farcs743, 3k$\times$2k CCD).  All data are freely available in the LCO Science Archive.\footnote{\url{https://archive.lco.global/}}  IAU observatory codes for all observations are listed in the observation table.   The data set covers 2018 December 15 to 2018 December 26 UTC.  Exposure times were 60 to 150~s per image, totaling 200 to 400~s per filter at each observational epoch.  The LCOGT pipeline \citep{mccully18-banzai} produces a raw photometry catalog for sources in each image.   We used this catalog to derive calibration zero-points and color corrections after cross-matching to the Pan-STARRS~1 DR1 catalog with the \texttt{calviacat} software \citep{kelley19-calviacat}.  The small amount of trailing $\leq1$\arcsec{} did not significantly affect the calibration, however the star FWHM measurements are slightly affected.  The median ellipticity of the stars was 0.12, and the maximum was 0.23.  We adopt a minimum uncertainty of 0.02~mag for this method.  Photometry of the comet was measured in 10,000~km radius apertures.

  \subsection{TRAPPIST Telescopes}

  We observed comet 243P with both TRAPPIST 60-cm robotic telescopes starting 4 days after the outburst, over the time period 2018 December 15 to 2019 February 26 UTC when the comet was at 2.72~au from the Sun and at 2.96~au from the Earth.  As indicated in Table~\ref{tab:obs}, we observed the comet 6 nights with TRAPPIST-North (TN) located at Oukaimeden Observatory, Morocco and 2 nights with TRAPPIST-South (TS) located at ESO/La Silla Observatory, Chile \citep{jehin11-trappist}.  The observations used the Johnson-Cousins $B$, $V$, $R$ and $I$ filters and an exposure time of 120s at the beginning of the observations with 240s at the end when the comet was fainter. We binned the pixels 2-by-2 and obtained a resulting plate scale of 1\farcs2 per pixel for TN and 1\farcs3 per pixel for TS. Standard procedures were used to calibrate the data, including flat-field correction with twilight sky \citep{moulane18-hmp-tgk}.  We derived the apparent magnitude in a 6\farcs3 radius apertures using the zero points and the extinction coefficients corresponding to each site. The resulting $R$-band photometry was converted to the PS1 system using the work of \citet{tonry12-ps1}.

  \subsection{Xingming Observatory}

  We observed the comet using the 60-cm Ningbo bureau of Education and Xinjiang observatory Telescope (NEXT) at Xingming Observatory on 2018 December 15 UTC.  Images were obtained through the Johnson $V$ and Cousins $R$ filters, at a scale of 0\farcs6 per pixel, flat-field calibrated with twilight sky frames.  Exposure times were 60~s, and the total integration times for each filter were 300~s for $V$ and 600~s for $R$. Color-corrected photometry was then derived using the Pan-STARRS~1 DR1 catalog, assuming a solar-colored comet.  We used an aperture of 7\farcs2 in radius, i.e., 10,000~km at the comet.  The telescope tracked sky at the sidereal rate, but the short exposure times prevented significant trailing of the comet.

  \subsection{Lulin Observatory}

  The photometric observations of this comet were performed on the nights of 2018 December 19 and 20 UTC using Johnson $V$ and Cousins $R$ filters with the Lulin One-meter Telescope (LOT) on the Lulin Observatory of the National Central University in Taiwan. The camera was operated with a Princeton Instruments Sophia 2048$\times$2048 pixel CCD, which provides a field of view of $13\farcm2\times 13\farcm2$ field of view at a spatial resolution of 0\farcs39~pixel$^{-1}$. The exposure times were 180~s for both $V$ and $R$ filters, which trailed the stars a small amount (1\arcsec).  Data calibration included the use of twilight sky flats.  An aperture a radius of 7\arcsec{} centered at the nucleus was used to determine the magnitude of 243P.  The data were calibrated using the Pan-STARRS~1 DR1 catalog.

  \subsection{Lowell Observatory}

  We observed the comet with the 4.3-m Lowell Discovery Telescope (LDT) on 2018 December 16 and 2019 January 12 UTC.  On both nights we used the Large Monolithic Imager \citep[LMI;][]{massey13-lmi}, a 6k$\times$6k CCD with a pixel scale of 0\farcs12.  The CCD was read out in 2$\times$2 binning mode (0\farcs24~pix$^{-1}$).  On December 16 we observed with the SDSS-$r'$ filter, and the narrow-band Hale-Bopp dust continuum filters $BC$ and $RC$ (0.445 and 0.713~\micron, 1\% bandwidth) \citep{farnham00}.  On January 12 we only used the SDSS-$r$ filter.  On both nights, data calibration included the use of twilight sky flats.  The $r'$ filter was calibrated to the PS1 system using background stars in the same manner as the LCOGT images.  The stars in the 2018 December 16 image were trailed 1\farcs6, but this was accounted for with increased aperture sizes and the image quality was measured perpendicular to the trailing.  The $BC$ and $RC$ data were calibrated using standard stars HD~6815, 31331, and 37112 observed over a wide range of airmasses.  The comet's brightness was measured within a 10,000~km aperture radius.

  \subsection{Apache Point Observatory}

  The comet was observed with the Apache Point Observatory's Astrophysical Consortium (ARC) 3.5-m telescope on 2018 December 15 and 2018 December 31 UTC. The Nasmyth-mounted guide camera, a 300$\times$300 pixel CCD with a pixel scale of 0\farcs41, equipped with Johnson and Cousins filters, was used for observations during both of these nights.  Images were taken through the $V$- and/or $R$-band filters, and the data calibration included the use of dome flat field images.  The data were calibrated to the PS1 system using (untrailed) background stars, and the comet's brightness measured within a 10,000~km aperture radius.

  \subsection{Catalina Sky Survey}

  A personal inquiry to the Catalina Sky Survey (CSS)\footnote{\url{http://www.lpl.arizona.edu/css/}} determined that comet 243P had been covered by the survey around the time of the outburst.  CSS operates two survey telescopes in the Santa Catalina mountains north of Tucson, AZ.  The comet was observed in one survey field on two nights (2018 December 10 and 14 UTC) by the 0.7-m Schmidt telescope on Mt.\ Bigelow (MPC site code 703) and in one field on one night (2018 December 14 UTC) by the 1.5-m Cassegrain reflector on Mt.\ Bigelow (MPC site code G96). Each survey field consists of four 30~sec exposures and no filter was employed as these were NEO survey observations.  Both telescopes are equipped with 10,560$\times$10,560 pixel CCD cameras which are operated binned $2\times2$, giving pixel scales of 3\farcs00 and 1\farcs52 for 703 and G96 respectively.  The data processing pipeline is described by \citet{seaman22-css-archive}.  Although the telescope tracked the sky at the sidereal rate, the comet was essentially untrailed (motion $\leq0.1$~pix).

  Source catalogs were extracted for each frame using Source Extractor \citep{bertin96} with aperture sizes scaled to correspond to 10,000\,km at the distance of the comet at the time of observations. Zero-points were determined through a cross-match to the PS1 system using background stars (see LCOGT calibration) within a 0.5\degr\ radius of the comet position since the field of view is larger (many degrees) than can be supported in a single PS1 catalog query. The comet observations were color-corrected using the $g-r$ color of the background stars and the derived color of the comet.

\end{CJK*}
\bibliographystyle{aasjournal}
\bibliography{apj-jour,references,extra}

\end{document}